\documentclass[%
 reprint,
superscriptaddress,
 amsmath,amssymb,
 aps,prl,
floatfix,
]{revtex4-2}

\usepackage{graphicx}
\usepackage{dcolumn}
\usepackage{bm}
\usepackage{braket}
\usepackage{hyperref}
\hypersetup{pdfpagemode=FullScreen,
colorlinks=true, citecolor = blue, linkcolor=blue, filecolor=blue}
\usepackage[mathlines]{lineno}
\usepackage{amsmath}
\usepackage[T1]{fontenc}
\usepackage{color}

\usepackage[normalem]{ulem}

\begin{document}

\preprint{APS/123-QED}

\title{On-chip Radio Frequency Maser}

\author{Hongliang Wu}%
\affiliation{%
Key Laboratory of Advanced Optoelectronic Quantum Architecture and Measurements of Ministry of Education, Center for Interdisciplinary Science of Optical Quantum and NEMS Integration, School of Physics, Beijing Institute of Technology, Beijing 100081, China
}%
\affiliation{%
School of Integrated Circuits and Electronics, Beijing Institute of Technology, Beijing 100081, China
}%
\affiliation{%
Center for Photonic Quantum Precision Measurement, Advanced Research Institute of Multidisciplinary Science, Beijing Institute of Technology, Beijing 100081, China
}%

\author{Zhengtao Wang}%
\affiliation{%
Key Laboratory of Advanced Optoelectronic Quantum Architecture and Measurements of Ministry of Education, Center for Interdisciplinary Science of Optical Quantum and NEMS Integration, School of Physics, Beijing Institute of Technology, Beijing 100081, China
}%
\affiliation{%
Center for Photonic Quantum Precision Measurement, Advanced Research Institute of Multidisciplinary Science, Beijing Institute of Technology, Beijing 100081, China
}%

\author{Yuchen Han}%
\affiliation{%
Key Laboratory of Advanced Optoelectronic Quantum Architecture and Measurements of Ministry of Education, Center for Interdisciplinary Science of Optical Quantum and NEMS Integration, School of Physics, Beijing Institute of Technology, Beijing 100081, China
}%
\affiliation{%
Center for Photonic Quantum Precision Measurement, Advanced Research Institute of Multidisciplinary Science, Beijing Institute of Technology, Beijing 100081, China
}%

\author{Liu Yang}
\affiliation{%
Key Laboratory of Advanced Optoelectronic Quantum Architecture and Measurements of Ministry of Education, Center for Interdisciplinary Science of Optical Quantum and NEMS Integration, School of Physics, Beijing Institute of Technology, Beijing 100081,  China
}%

\author{Zhiwei Wang}
\affiliation{%
Key Laboratory of Advanced Optoelectronic Quantum Architecture and Measurements of Ministry of Education, Center for Interdisciplinary Science of Optical Quantum and NEMS Integration, School of Physics, Beijing Institute of Technology, Beijing 100081, China
}%
\affiliation{%
Beijing Key Lab of Nanophotonics and Ultrafine Optoelectronic Systems, Beijing Institute of Technology, Beijing 100081, China.
}%
\affiliation{%
International Centre for Quantum Materials, Beijing Institute of Technology, Zhuhai 519000, China.
}%

\author{Yeliang Wang}
\affiliation{%
School of Integrated Circuits and Electronics, Beijing Institute of Technology, Beijing 100081, China
}%
\affiliation{%
MIIT Key Laboratory for Low-Dimensional Quantum Structure and Devices, Beijing Institute of Technology, Beijing 100081, China
}%

\author{Dezhi Zheng}
\affiliation{%
State Key Laboratory of Environment Characteristics and Effects for Near-space, Beijing Institute of Technology, Beijing 100081, China
}%
\affiliation{%
MIIT Key Laboratory of Complex-field Intelligent Sensing, Beijing Institute of Technology, Beijing 100081, China
}%
\affiliation{%
State Key Laboratory of CNS/ATM, Beijing Institute of Technology, Beijing 100081, China
}%

\author{Bo Zhang}
\email{bozhang_quantum@bit.edu.cn}
\affiliation{%
Key Laboratory of Advanced Optoelectronic Quantum Architecture and Measurements of Ministry of Education, Center for Interdisciplinary Science of Optical Quantum and NEMS Integration, School of Physics, Beijing Institute of Technology, Beijing 100081, China
}%
\affiliation{%
Center for Photonic Quantum Precision Measurement, Advanced Research Institute of Multidisciplinary Science, Beijing Institute of Technology, Beijing 100081, China
}%
\affiliation{%
State Key Laboratory of Environment Characteristics and Effects for Near-space, Beijing Institute of Technology, Beijing 100081, China
}%
\affiliation{%
MIIT Key Laboratory of Complex-field Intelligent Sensing, Beijing Institute of Technology, Beijing 100081, China
}%

\author{Jun Zhang}
\email{zhjun@bit.edu.cn}
\affiliation{%
State Key Laboratory of Environment Characteristics and Effects for Near-space, Beijing Institute of Technology, Beijing 100081, China
}%
\affiliation{%
MIIT Key Laboratory of Complex-field Intelligent Sensing, Beijing Institute of Technology, Beijing 100081, China
}%
\affiliation{%
State Key Laboratory of CNS/ATM, Beijing Institute of Technology, Beijing 100081, China
}%

\begin{abstract}
Room-temperature solid-state masers offer exceptional frequency selectivity and ultra-low noise for weak-signal detection. However, their reliance on bulky metallic resonators has significantly hindered integration, miniaturization, and extension to lower frequencies. Here, we demonstrate the first on-chip radio-frequency maser operating at room temperature, exploiting optically pumped triplet states of pentacene. The device produces stimulated emission at 106.62 MHz and enables ultra-sensitive microwave magnetic-field detection with a sensitivity of ($\sim 10\,\rm{fT/\sqrt{Hz}}$), functioning simultaneously as a local oscillator and a sensor. By actively controlling microwave dissipation, we achieve efficient regulation of the maser output, revealing a key mechanism for tuning emission in open cavity-free systems. This work extends pentacene-based masers into the radio-frequency regime and establishes a highly integrated on-chip architecture for room-temperature masers, offering a new pathway toward portable quantum devices.
\end{abstract}

\maketitle

\textbf{Introduction-} Pentacene doped $p$-terphenyl \cite{Oxborrow2012} and nitrogen‑vacancy (NV) centers in diamond \cite{Breeze2018} are established room‑temperature solid‑state masers platform \cite{PhysRev.99.1264}. Optical pumping provides high spin polarization and long coherence without cryogenics, enabling coherent sources \cite{Breeze2017,long2026lbandmilliwattroomtemperaturesolidstate}, high‑gain microwave amplifiers \cite{https://doi.org/10.1002/advs.202401904,10.1063/5.0271776,doi:10.1126/sciadv.ade6527,PhysRevX.14.041066}, and ultra‑weak signal detectors \cite{wu2025detectingaxiondarkmatter,doi:10.1126/sciadv.ade1613}.

However, existing masers rely on bulky metallic resonators. High‑permittivity dielectrics \cite{Breeze2015} reduce the magnetic mode volume ($V_{\rm{m}}$) while maintaining a relatively high loaded quality factor ($Q_{\rm{L}}$), but remain too large for on-chip integration and grow still larger at lower frequencies because resonator size scales inversely with frequency.

Pentacene is attractive for pulsed masers because of its high transient polarization and spin density \cite{mr-2-33-2021}, chemical tunability \cite{Bogatko2016,adma.202300441,Schroder2022,KOUSKOV19959,Kohler1993,Attwood2023} and scalable preparation \cite{Cui2020}. Previous pentacene masers operated near 1.45 GHz and, under Zeeman tuning, near 9.4 GHz \cite{https://doi.org/10.1002/advs.202401904}. Because Zeeman tuning raises rather than lowers the transition frequency, the sub-GHz (VHF/UHF) regime remains largely unexplored, despite its relevance to underground and underwater detection \cite{doi:10.1049/ecej:19960402}, low‑frequency radar \cite{6967722}, and astronomical observation \cite{refId0}.

Here, we realize the first on‑chip pentacene low‑frequency (LF) maser using the previously unused optically polarized triplet subspace and a compact on‑chip resonator. At room temperature and zero field, active control of microwave dissipation switches the 106.6 MHz device between coherent oscillation and magnetic sensing and reveals the dependence of the maser threshold on $Q_{\rm{L}}$.


\begin{figure}[htbp!]
\centering
\includegraphics[width=0.5\textwidth]{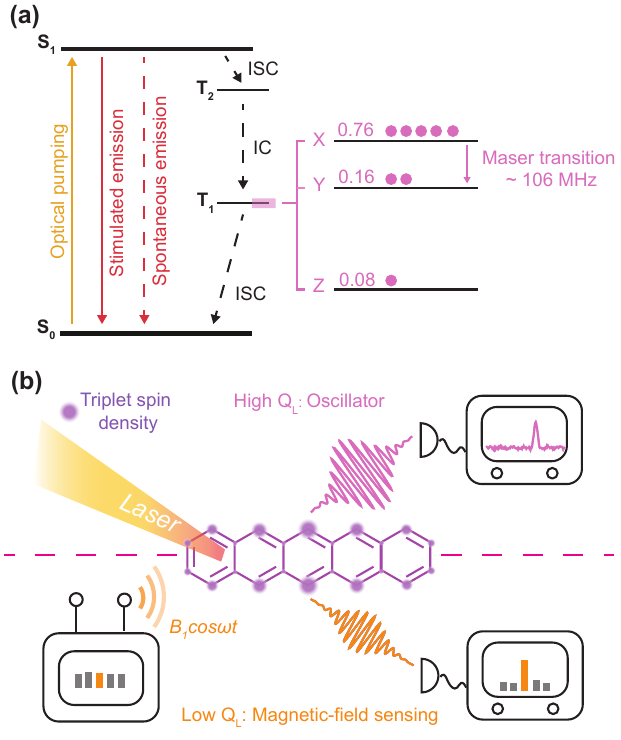}
\caption{\label{principle}
\textbf{The mechanism of LF maser.} (a) Pentacene molecules are pumped by a 590-nm pulse, followed by ISC and rapid IC to the first triplet excited state $\rm{T}_{1}$. In the absence of an external magnetic field, the $\rm{T}_{1}$ state spontaneously splits into the X, Y, and Z sublevels. The population ratio between the X and Y sublevels is approximately 0.76: 0.16. X-Y inversion drives the approximately 106-MHz transition. (b) High $Q_{\rm{L}}$ supports self-oscillation while low $Q_{\rm{L}}$ suppresses oscillation and enables magnetic-field sensing.
}
\end{figure}

\textbf{Principle of LF maser-} As shown in Fig. \ref{principle}(a), Optical excitation of 0.1\% pentacene doped p-terphenyl at 590 nm transfers approximately 62.5\% \cite{10.1063/1.1499124} of $\rm{S}_{1}$ molecules to $\rm{T}_{2}$ through spin-selective intersystem crossing (ISC), producing X:Y:Z populations of 0.76:0.16:0.08 \cite{10.1063/1.442520}. Rapid internal conversion (IC) to $T_1$ preserves this distribution, and sublevel-dependent relaxation maintains the inversion required for room-temperature masing.

Existing pentacene masers exploit the X-Z sublevel pair, which exhibits the largest population difference and typically operates around 1.4495 GHz using dielectric $\rm{TE}_{01\delta}$ resonators. In this work, we instead use the X‑Y sublevels. Their energy splitting is approximately 106 MHz \cite{10.1063/1.1326069}, placing the maser in the radio‑frequency (VHF) band. Although the X‑Y population inversion is smaller than that of X‑Z, the X‑Y transition offers two decisive physical advantages for low‑frequency operation. First, the Y sublevel has a shorter lifetime than the Z sublevel, and the spin–lattice relaxation rate between X and Y is lower than that between X and Z \cite{Wu2019}. This reduces the effective decay rate of the inversion. Second, the lower operating frequency ($\sim$ 106 MHz) relaxes the requirement on the $Q_{\rm{L}}$, since the cavity-photon decay rate $\kappa_{\rm{c}}$ is given by $\kappa_{\rm{c}}=freq./Q_{\rm{L}}$. Together, these features establish the previously overlooked X–Y subspace as a viable and efficient gain channel for realizing a compact, on-chip radio-frequency maser.

Figure \ref{principle}(b) demonstrates the two operating regimes of the LF maser. At a high $Q_{\rm{L}}$, the reduced photon dissipation rate allows the system to enter the oscillator regime, producing intense pulsed emissions. As the $Q_{\rm{L}}$ is lowered, self-oscillation is suppressed and the system transitions into the sensor regime, where it responds with high sensitivity to externally applied magnetic signals.


\begin{figure*}[htbp!]
\centering
\includegraphics[width=1\textwidth]{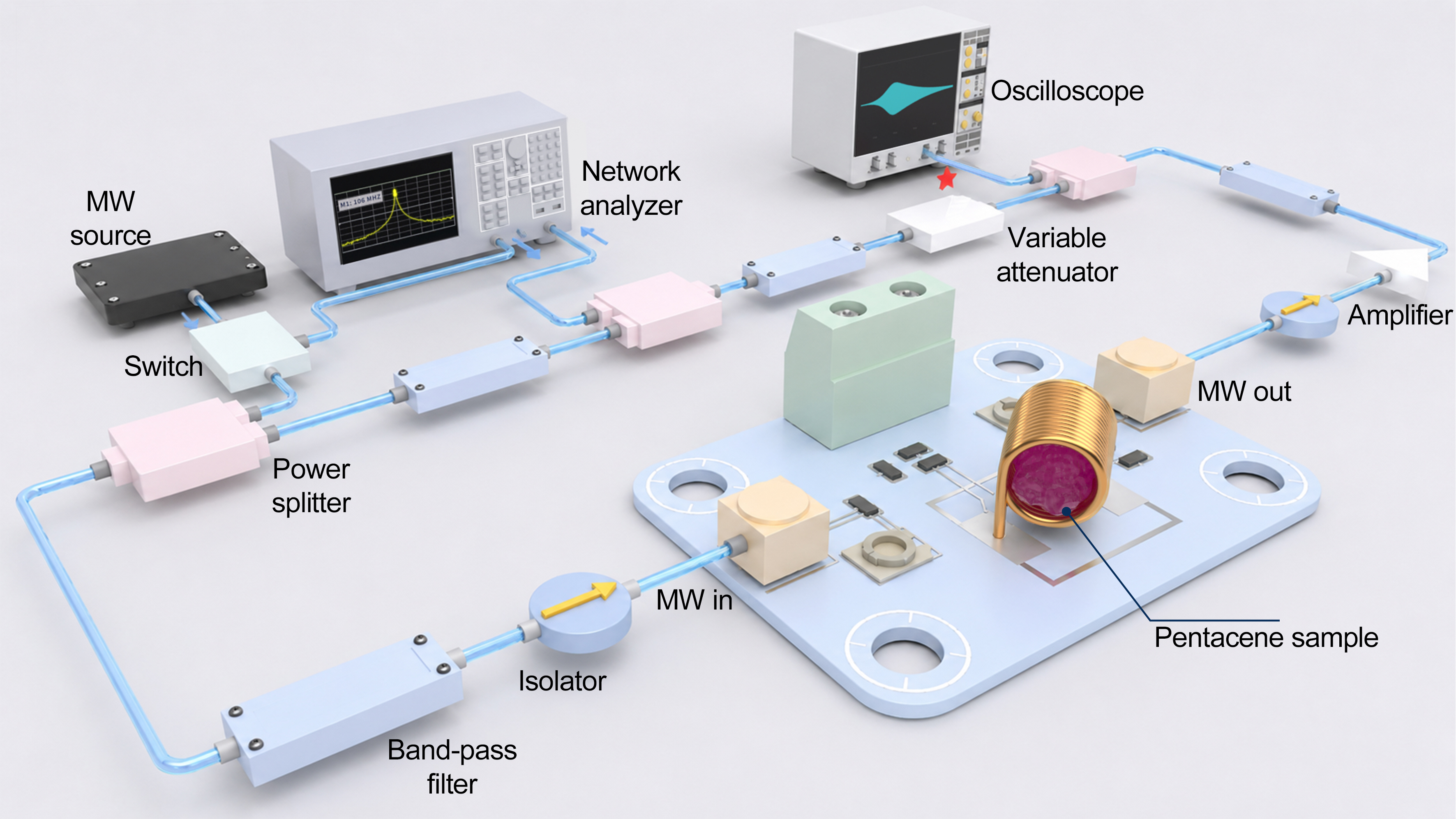}
\caption{\label{LC circuit and feedback}
\textbf{Setup for the LF maser.} A gain-adjustable feedback loop tunes the LC circuit $Q_{\rm{L}}$ and switches between oscillator and sensing regimes. A logarithmic detector can be added at the position marked by a red star to convert the time-domain signal into an envelope signal.
}
\end{figure*}

\textbf{Design of the on-chip resonator-} To realize stimulated emission on the X‑Y transition, we replace the conventional metallic‑cavity or dielectric resonator with a compact on‑chip inductor-capacitor (LC) resonant circuit (see Supplemental Materials). As shown in Fig. \ref{LC circuit and feedback}, The pentacene sample is placed directly inside the hollow inductor. The resonant magnetic field is aligned with the inductor axis and is strongly concentrated within the solenoid region, with only minimal leakage at the ends. From finite-element simulations, the effective magnetic mode volume is $V_{\rm{m}}=\int \frac{|\rm{H(r)}|^{2}}{|\rm{H}_{\rm{max}}(\rm{r})|^{2}} \rm{d}V$=37.6\,$\rm{mm}^{3}$, only slightly larger than the solenoid’s physical volume. This small $V_{\rm{m}}$ enhances the spin‑field coupling (via the Purcell factor), making the X‑Y transition viable despite its moderate polarization difference.

To achieve both quantum oscillation and high‑sensitivity magnetic‑field sensing, the LC circuit is connected to a feedback loop, as depicted in Fig. \ref{LC circuit and feedback}. The variable-attenuation feedback loop tunes $Q_{\rm{L}}$ from its intrinsic value of approximately 40 to several thousand. For fixed spin polarization, $Q_{\rm{L}}$ above threshold gives coherent oscillation, whereas $Q_{\rm{L}}$ below threshold suppresses self-oscillation and enables signal amplification. We keep $Q_{\rm{L}}$ below approximately $3\times10^3$ to avoid classical loop oscillation. This feedback‑based $Q_{\rm{L}}$ control provides a versatile platform for cavity quantum electrodynamics with solid‑state spins \cite{Ng2021}, magnons \cite{Yao2017,PhysRevLett.130.146702}, and atomic vapors \cite{doi:10.1126/sciadv.abe0719} at room temperature.

As shown in Fig. \ref{LC circuit and feedback}, two power dividers are employed to couple a vector network analyzer (VNA) to the circuit for $S$-parameter measurements. The $Q_{\rm{L}}$ is obtained from the measured transmission coefficient $S_{\rm{21}}$. A switch is placed at the input port to enable switching between the VNA and a microwave source, allowing subsequent magnetic-field sensing. The overall output of the system is taken from another power divider, which is connected to an oscilloscope to display the final maser output signal.


\begin{figure}[htbp!]
\centering
\includegraphics[width=0.5\textwidth]{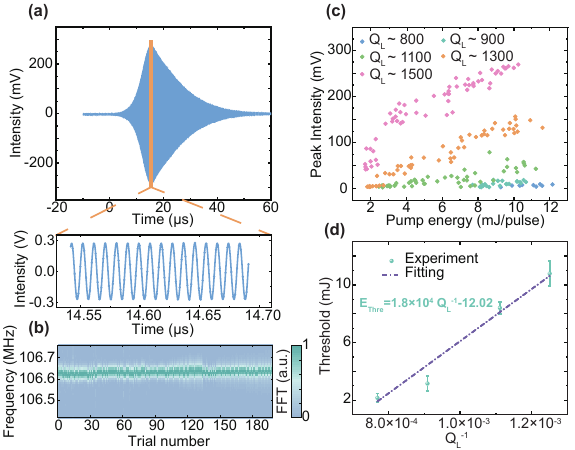}
\caption{\label{maser}
\textbf{Performance of the microwave quantum oscillator.} (a) The upper panel shows the time-domain data of LF maser pulses. The lower panel provides a detailed view of the position of maximum maser output. The dots represent the data points acquired by the oscilloscope, and the solid line is the sinusoidal fit. (b) Emission spectra of the LF maser. Emission remains centered at 106.62 MHz over repeated trials. (c) The maser output peak intensity versus the pump energy under different $Q_{\rm{L}}$. The relationship clearly follows a square-root function. (d) Linear relationship between the $Q_{\rm{L}}^{-1}$ and the threshold pump energy. The threshold is defined at 1 $\mu \rm{W}$ maser output.}
\end{figure}

\textbf{Microwave quantum oscillator-} Figure \ref{maser}(a) shows the typical time‑domain output of the LF maser operating at the resonance frequency of 106.62 MHz (which matches the X‑Y sublevel splitting; see Supplemental Material) under a pump energy of 9 mJ/pulse and an enhanced loaded quality factor $Q_{\rm{L}} \approx 1500$. The output signal exhibits a clear delay relative to the optical pump pulse. This delay arises from two physical contributions: the build-up time required for the microwave field to exceed the maser threshold, and the ring‑up time of the resonator, $\tau_{\rm{R}}=\frac{2 Q_{\rm{L}}}{\rm{\omega}_{\rm{c}}} \approx 4.5\,\mu\rm{s}$. The latter is a direct consequence of the cavity’s energy storage time, which sets the response speed of the oscillator.

Figure \ref{maser}(b) displays fast Fourier transform (FFT) spectra acquired over multiple experimental runs at a pump repetition rate of 10 Hz. The spectra show excellent frequency stability, confirming that the maser oscillation is locked to the X‑Y transition. Unlike dielectric resonators, which are susceptible to laser‑induced heating and frequency shifts \cite{Xiang2025,DielectricProperties}, the LC circuit is essentially insensitive to such heating. Provided the pump energy remains below the damage thresholds of the pentacene crystal and the inductor, the output frequency remains stable, a key advantage for practical operation.

To systematically characterize the role of $Q_{\rm{L}}$ in the maser dynamics, we measured the output peak intensity as a function of pump energy for several fixed $Q_{\rm{L}}$ values (Fig. \ref{maser}(c)). For a given $Q_{\rm{L}}$, the intensity scales as the square root of the pump energy. This dependence follows from two basic relations: the detected electrical signal intensity is proportional to the square root of the output microwave power ($\rm{Intensity} \propto \sqrt{\rm{Power}}$), while the output power is proportional to the number of polarized spins $N$ excited by the pump, and $N$ is itself proportional to the pump energy \cite{Breeze2017}. Hence $\rm{Intensity} \propto \sqrt{E_{\rm{pump}}}$. This square‑root law directly reflects the maser’s operating point above threshold, where the output power grows linearly with the inversion.

As shown in Fig. \ref{maser}(d), The threshold pump energy decreases inversely with $Q_{\rm{L}}$, consistent with theoretical predictions \cite{Breeze2015,Breeze2018,Zollitsch2023}. Extrapolating the fit to the intrinsic $Q_{\rm{L}} \sim 40$ yields a threshold pump energy of about 348 mJ, which far exceeds the damage threshold of pentacene. This explains why the feedback‑enhanced $Q_{\rm{L}}$ is essential for observing LF maser oscillation. It also indicates that improving the intrinsic $Q_{\rm{L}}$ of the on‑chip resonator remains an important objective.


\textbf{Magnetic-field sensing-} By reducing the $Q_{\rm{L}}$ to about 400, the self-oscillation is suppressed. In this regime, the device emits only in response to an injected signal, converting the system into a narrow-band weak-signal sensor. To optimize sensitivity, we increase the pump energy to the maximum nondamaging level ($\sim 9$ mJ). We set the frequency of the microwave source to match both the resonator and the spin transition frequency, $\omega=2\pi \times 106.62\,\rm{MHz}$, and steped by 3-dB. A logarithmic detector was placed at the position marked by the red star in Fig. \ref{LC circuit and feedback} to records the output envelope, enabling precise evaluation of the sensitivity. 

To relate the applied microwave power $P_{\rm{MW}}$ to the actual magnetic field $B_{1}$ experienced by the pentacene sample, we use the standard resonator‑based conversion factor $C=B_{1}/\sqrt{P_{\rm{MW}}}=\sqrt{\frac{2 \mu_{0} Q_{\rm{L}}}{V_{\rm{m}} \omega}}$ \cite{poole1983electron}, where $\mu_{0}=4\pi \times 10^{-7}\,\rm{H/m}$ is the vacuum permeability. For our system, $C=6.32\,\rm{mT/\sqrt{W}}$.

\begin{figure}[htbp!]
\centering
\includegraphics[width=0.5\textwidth]{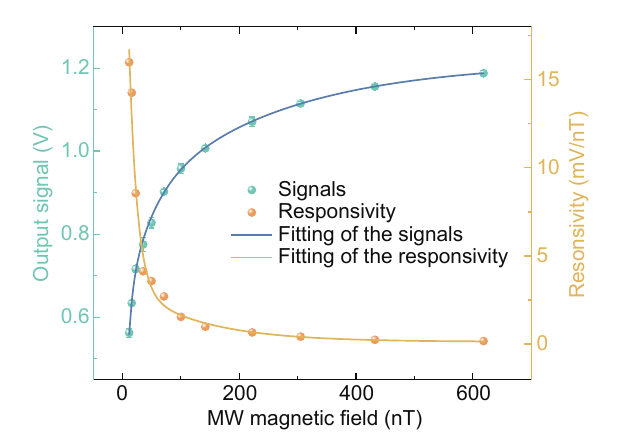}
\caption{\label{detection}
\textbf{The sensitivity of the magnetic-field sensing.} The green points represent the signal strengths measured at different microwave magnetic fields and are fitted by an exponential equation. The slope at each data point can be obtained from the fitted curve, which is the responsivity of the system to the magnetic field.}
\end{figure}

The measured maser intensity $S$ versus $B_{1}$ is shown as green dots in Fig. \ref{detection}, exhibiting a clear nonlinear response. The magnetic‑field responsivity is defined as $R=\frac{\partial S}{\partial B_{1}}$. We performed a nonlinear fit to the data and computed the derivative, yielding the yellow dots in Fig. \ref{detection}. The sensitivity (noise‑equivalent magnetic field) is then $\eta=\frac{\sigma_{\rm{s}}}{R\sqrt{2 \Delta f}}$, where $\sigma_{\rm{s}}=5.24\,\rm{mV}$ is the single‑measurement uncertainty and $\Delta f=500\,\rm{MHz}$ is the acquisition bandwidth of the oscilloscope. The logarithmic detector ensures that noise across the entire bandwidth contributes proportionally, providing a flat noise floor. The optimal sensitivity achieved is $\eta=10.38\,\rm{fT/\sqrt{Hz}}$.

Although this value is slightly lower than that reported for pentacene masers using the X‑Z transition \cite{10.1063/5.0271776,wu2025detectingaxiondarkmatter}, it nevertheless represents several order‑of‑magnitude improvement over state‑of‑the‑art microwave magnetic‑field sensors based on NV centers \cite{doi:10.1126/sciadv.abq8158,Meinel2021,PhysRevApplied.19.054095,Wang2015,Chen2023,PhysRevApplied.12.044039,Eisenach2021}. This demonstrates that the on‑chip LF maser is a competitive platform for sensitive microwave magnetic-field detection, with the added advantages of compactness and room‑temperature operation.

\textbf{Discussion-} We demonstrate the first room-temperature, zero-field on-chip RF maser based on pentacene. Active $Q_{\rm{L}}$ control switches the device between a frequency-stable 106.62 MHz oscillator and a high‑sensitivity microwave magnetometer reaching approximately $\sim 10\,\rm{fT/\sqrt{Hz}}$. The sensitivity is slightly lower than that of X-Z pentacene masers but several orders of magnitude better than that of NV-center microwave magnetometers.

The LC resonator and feedback-loop electronics are compatible with printed circuit board (PCB) technology, allowing the entire system to be integrated onto a single PCB platform. Moreover, the operating frequency can be flexibly tuned by modifying the inductance and capacitance of the LC resonator, enabling extension to other maser regimes. Our approach overcomes the long‑standing limitation of conventional room‑temperature solid‑state masers-their reliance on bulky metallic resonators-and opens a new route toward portable, chip‑scale quantum devices. Beyond providing a versatile platform for studying cavity quantum electrodynamics (cQED) under ambient conditions, this work advances the practical implementation of maser‑based quantum precision measurement technologies such as nuclear magnetic resonance (NMR) \cite{https://doi.org/10.1002/cphc.201901056,Suefke2017,https://doi.org/10.1002/anie.202525699,https://doi.org/10.1002/mrm.21396}, portable microwave oscillators and high‑sensitivity magnetometry \cite{10.1063/5.0181318,PhysRevApplied.23.054064}.

\vspace{8pt} %
\textbf{Acknowledgments-} We sincerely thank Hao Wu for the initial inspiration of this work and for the stimulating discussions. This study was supported by NSF of China (Grant No.\,T2522007, No.\,12441502, No.\,12374462, No.\,12321004).

\textbf{Data availability-} The data that support the findings of this article are not publicly available. The data are available from the authors upon reasonable request.

\nocite{*}


\begin{thebibliography}{48}%
\makeatletter
\providecommand \@ifxundefined [1]{%
 \@ifx{#1\undefined}
}%
\providecommand \@ifnum [1]{%
 \ifnum #1\expandafter \@firstoftwo
 \else \expandafter \@secondoftwo
 \fi
}%
\providecommand \@ifx [1]{%
 \ifx #1\expandafter \@firstoftwo
 \else \expandafter \@secondoftwo
 \fi
}%
\providecommand \natexlab [1]{#1}%
\providecommand \enquote  [1]{``#1''}%
\providecommand \bibnamefont  [1]{#1}%
\providecommand \bibfnamefont [1]{#1}%
\providecommand \citenamefont [1]{#1}%
\providecommand \href@noop [0]{\@secondoftwo}%
\providecommand \href [0]{\begingroup \@sanitize@url \@href}%
\providecommand \@href[1]{\@@startlink{#1}\@@href}%
\providecommand \@@href[1]{\endgroup#1\@@endlink}%
\providecommand \@sanitize@url [0]{\catcode `\\12\catcode `\$12\catcode
  `\&12\catcode `\#12\catcode `\^12\catcode `\_12\catcode `\%12\relax}%
\providecommand \@@startlink[1]{}%
\providecommand \@@endlink[0]{}%
\providecommand \url  [0]{\begingroup\@sanitize@url \@url }%
\providecommand \@url [1]{\endgroup\@href {#1}{\urlprefix }}%
\providecommand \urlprefix  [0]{URL }%
\providecommand \Eprint [0]{\href }%
\providecommand \doibase [0]{https://doi.org/}%
\providecommand \selectlanguage [0]{\@gobble}%
\providecommand \bibinfo  [0]{\@secondoftwo}%
\providecommand \bibfield  [0]{\@secondoftwo}%
\providecommand \translation [1]{[#1]}%
\providecommand \BibitemOpen [0]{}%
\providecommand \bibitemStop [0]{}%
\providecommand \bibitemNoStop [0]{.\EOS\space}%
\providecommand \EOS [0]{\spacefactor3000\relax}%
\providecommand \BibitemShut  [1]{\csname bibitem#1\endcsname}%
\let\auto@bib@innerbib\@empty
\bibitem [{\citenamefont {Oxborrow}\ \emph {et~al.}(2012)\citenamefont
  {Oxborrow}, \citenamefont {Breeze},\ and\ \citenamefont
  {Alford}}]{Oxborrow2012}%
  \BibitemOpen
  \bibfield  {author} {\bibinfo {author} {\bibfnamefont {M.}~\bibnamefont
  {Oxborrow}}, \bibinfo {author} {\bibfnamefont {J.~D.}\ \bibnamefont
  {Breeze}},\ and\ \bibinfo {author} {\bibfnamefont {N.~M.}\ \bibnamefont
  {Alford}},\ }\bibfield  {title} {\bibinfo {title} {Room-temperature
  solid-state maser},\ }\href {https://doi.org/10.1038/nature11339} {\bibfield
  {journal} {\bibinfo  {journal} {Nature}\ }\textbf {\bibinfo {volume} {488}},\
  \bibinfo {pages} {353} (\bibinfo {year} {2012})}\BibitemShut {NoStop}%
\bibitem [{\citenamefont {Breeze}\ \emph {et~al.}(2018)\citenamefont {Breeze},
  \citenamefont {Salvadori}, \citenamefont {Sathian}, \citenamefont {Alford},\
  and\ \citenamefont {Kay}}]{Breeze2018}%
  \BibitemOpen
  \bibfield  {author} {\bibinfo {author} {\bibfnamefont {J.~D.}\ \bibnamefont
  {Breeze}}, \bibinfo {author} {\bibfnamefont {E.}~\bibnamefont {Salvadori}},
  \bibinfo {author} {\bibfnamefont {J.}~\bibnamefont {Sathian}}, \bibinfo
  {author} {\bibfnamefont {N.~M.}\ \bibnamefont {Alford}},\ and\ \bibinfo
  {author} {\bibfnamefont {C.~W.~M.}\ \bibnamefont {Kay}},\ }\bibfield  {title}
  {\bibinfo {title} {Continuous-wave room-temperature diamond maser},\ }\href
  {https://doi.org/10.1038/nature25970} {\bibfield  {journal} {\bibinfo
  {journal} {Nature}\ }\textbf {\bibinfo {volume} {555}},\ \bibinfo {pages}
  {493} (\bibinfo {year} {2018})}\BibitemShut {NoStop}%
\bibitem [{\citenamefont {Gordon}\ \emph {et~al.}(1955)\citenamefont {Gordon},
  \citenamefont {Zeiger},\ and\ \citenamefont {Townes}}]{PhysRev.99.1264}%
  \BibitemOpen
  \bibfield  {author} {\bibinfo {author} {\bibfnamefont {J.~P.}\ \bibnamefont
  {Gordon}}, \bibinfo {author} {\bibfnamefont {H.~J.}\ \bibnamefont {Zeiger}},\
  and\ \bibinfo {author} {\bibfnamefont {C.~H.}\ \bibnamefont {Townes}},\
  }\bibfield  {title} {\bibinfo {title} {The maser---new type of microwave
  amplifier, frequency standard, and spectrometer},\ }\href
  {https://doi.org/10.1103/PhysRev.99.1264} {\bibfield  {journal} {\bibinfo
  {journal} {Phys. Rev.}\ }\textbf {\bibinfo {volume} {99}},\ \bibinfo {pages}
  {1264} (\bibinfo {year} {1955})}\BibitemShut {NoStop}%
\bibitem [{\citenamefont {Breeze}\ \emph {et~al.}(2017)\citenamefont {Breeze},
  \citenamefont {Salvadori}, \citenamefont {Sathian}, \citenamefont {Alford},\
  and\ \citenamefont {Kay}}]{Breeze2017}%
  \BibitemOpen
  \bibfield  {author} {\bibinfo {author} {\bibfnamefont {J.~D.}\ \bibnamefont
  {Breeze}}, \bibinfo {author} {\bibfnamefont {E.}~\bibnamefont {Salvadori}},
  \bibinfo {author} {\bibfnamefont {J.}~\bibnamefont {Sathian}}, \bibinfo
  {author} {\bibfnamefont {N.~M.}\ \bibnamefont {Alford}},\ and\ \bibinfo
  {author} {\bibfnamefont {C.~W.~M.}\ \bibnamefont {Kay}},\ }\bibfield  {title}
  {\bibinfo {title} {Room-temperature cavity quantum electrodynamics with
  strongly coupled dicke states},\ }\href
  {https://doi.org/10.1038/s41534-017-0041-3} {\bibfield  {journal} {\bibinfo
  {journal} {npj Quantum Information}\ }\textbf {\bibinfo {volume} {3}},\
  \bibinfo {pages} {40} (\bibinfo {year} {2017})}\BibitemShut {NoStop}%
\bibitem [{\citenamefont {Long}\ \emph {et~al.}(2026)\citenamefont {Long},
  \citenamefont {Attwood}, \citenamefont {Chan}, \citenamefont {Choubey},
  \citenamefont {Torun},\ and\ \citenamefont
  {Sathian}}]{long2026lbandmilliwattroomtemperaturesolidstate}%
  \BibitemOpen
  \bibfield  {author} {\bibinfo {author} {\bibfnamefont {S.}~\bibnamefont
  {Long}}, \bibinfo {author} {\bibfnamefont {M.}~\bibnamefont {Attwood}},
  \bibinfo {author} {\bibfnamefont {J.}~\bibnamefont {Chan}}, \bibinfo {author}
  {\bibfnamefont {P.}~\bibnamefont {Choubey}}, \bibinfo {author} {\bibfnamefont
  {H.}~\bibnamefont {Torun}},\ and\ \bibinfo {author} {\bibfnamefont
  {J.}~\bibnamefont {Sathian}},\ }\href {https://arxiv.org/abs/2511.09056}
  {\bibinfo {title} {L-band milliwatt room-temperature solid-state maser}}
  (\bibinfo {year} {2026}),\ \Eprint {https://arxiv.org/abs/2511.09056}
  {arXiv:2511.09056 [physics.app-ph]} \BibitemShut {NoStop}%
\bibitem [{\citenamefont {Wang}\ \emph {et~al.}(2024)\citenamefont {Wang},
  \citenamefont {Wu}, \citenamefont {Zhang}, \citenamefont {Yao}, \citenamefont
  {Zhang}, \citenamefont {Oxborrow},\ and\ \citenamefont
  {Zhao}}]{https://doi.org/10.1002/advs.202401904}%
  \BibitemOpen
  \bibfield  {author} {\bibinfo {author} {\bibfnamefont {K.}~\bibnamefont
  {Wang}}, \bibinfo {author} {\bibfnamefont {H.}~\bibnamefont {Wu}}, \bibinfo
  {author} {\bibfnamefont {B.}~\bibnamefont {Zhang}}, \bibinfo {author}
  {\bibfnamefont {X.}~\bibnamefont {Yao}}, \bibinfo {author} {\bibfnamefont
  {J.}~\bibnamefont {Zhang}}, \bibinfo {author} {\bibfnamefont
  {M.}~\bibnamefont {Oxborrow}},\ and\ \bibinfo {author} {\bibfnamefont
  {Q.}~\bibnamefont {Zhao}},\ }\bibfield  {title} {\bibinfo {title} {Tailoring
  coherent microwave emission from a solid-state hybrid system for
  room-temperature microwave quantum electronics},\ }\href
  {https://doi.org/https://doi.org/10.1002/advs.202401904} {\bibfield
  {journal} {\bibinfo  {journal} {Advanced Science}\ }\textbf {\bibinfo
  {volume} {11}},\ \bibinfo {pages} {2401904} (\bibinfo {year}
  {2024})}\BibitemShut {NoStop}%
\bibitem [{\citenamefont {Han}\ \emph {et~al.}(2026)\citenamefont {Han},
  \citenamefont {Wu}, \citenamefont {Wu}, \citenamefont {Oxborrow},
  \citenamefont {Li}, \citenamefont {Wang}, \citenamefont {Zheng},
  \citenamefont {Budker}, \citenamefont {Zhang},\ and\ \citenamefont
  {Zhang}}]{10.1063/5.0271776}%
  \BibitemOpen
  \bibfield  {author} {\bibinfo {author} {\bibfnamefont {Y.}~\bibnamefont
  {Han}}, \bibinfo {author} {\bibfnamefont {H.}~\bibnamefont {Wu}}, \bibinfo
  {author} {\bibfnamefont {H.}~\bibnamefont {Wu}}, \bibinfo {author}
  {\bibfnamefont {M.}~\bibnamefont {Oxborrow}}, \bibinfo {author}
  {\bibfnamefont {W.}~\bibnamefont {Li}}, \bibinfo {author} {\bibfnamefont
  {Y.}~\bibnamefont {Wang}}, \bibinfo {author} {\bibfnamefont {D.}~\bibnamefont
  {Zheng}}, \bibinfo {author} {\bibfnamefont {D.}~\bibnamefont {Budker}},
  \bibinfo {author} {\bibfnamefont {B.}~\bibnamefont {Zhang}},\ and\ \bibinfo
  {author} {\bibfnamefont {J.}~\bibnamefont {Zhang}},\ }\bibfield  {title}
  {\bibinfo {title} {Ultra-sensitive microwave magnetometry with
  organic-molecular sensors},\ }\href {https://doi.org/10.1063/5.0271776}
  {\bibfield  {journal} {\bibinfo  {journal} {Applied Physics Reviews}\
  }\textbf {\bibinfo {volume} {13}},\ \bibinfo {pages} {011403} (\bibinfo
  {year} {2026})}\BibitemShut {NoStop}%
\bibitem [{\citenamefont {Sherman}\ \emph {et~al.}(2022)\citenamefont
  {Sherman}, \citenamefont {Zgadzai}, \citenamefont {Koren}, \citenamefont
  {Peretz}, \citenamefont {Laster},\ and\ \citenamefont
  {Blank}}]{doi:10.1126/sciadv.ade6527}%
  \BibitemOpen
  \bibfield  {author} {\bibinfo {author} {\bibfnamefont {A.}~\bibnamefont
  {Sherman}}, \bibinfo {author} {\bibfnamefont {O.}~\bibnamefont {Zgadzai}},
  \bibinfo {author} {\bibfnamefont {B.}~\bibnamefont {Koren}}, \bibinfo
  {author} {\bibfnamefont {I.}~\bibnamefont {Peretz}}, \bibinfo {author}
  {\bibfnamefont {E.}~\bibnamefont {Laster}},\ and\ \bibinfo {author}
  {\bibfnamefont {A.}~\bibnamefont {Blank}},\ }\bibfield  {title} {\bibinfo
  {title} {Diamond-based microwave quantum amplifier},\ }\href
  {https://doi.org/10.1126/sciadv.ade6527} {\bibfield  {journal} {\bibinfo
  {journal} {Science Advances}\ }\textbf {\bibinfo {volume} {8}},\ \bibinfo
  {pages} {eade6527} (\bibinfo {year} {2022})}\BibitemShut {NoStop}%
\bibitem [{\citenamefont {Day}\ \emph {et~al.}(2024)\citenamefont {Day},
  \citenamefont {Isarov}, \citenamefont {Pappas}, \citenamefont {Johnson},
  \citenamefont {Abe}, \citenamefont {Ohshima}, \citenamefont {McCamey},
  \citenamefont {Laucht},\ and\ \citenamefont {Pla}}]{PhysRevX.14.041066}%
  \BibitemOpen
  \bibfield  {author} {\bibinfo {author} {\bibfnamefont {T.}~\bibnamefont
  {Day}}, \bibinfo {author} {\bibfnamefont {M.}~\bibnamefont {Isarov}},
  \bibinfo {author} {\bibfnamefont {W.~J.}\ \bibnamefont {Pappas}}, \bibinfo
  {author} {\bibfnamefont {B.~C.}\ \bibnamefont {Johnson}}, \bibinfo {author}
  {\bibfnamefont {H.}~\bibnamefont {Abe}}, \bibinfo {author} {\bibfnamefont
  {T.}~\bibnamefont {Ohshima}}, \bibinfo {author} {\bibfnamefont {D.~R.}\
  \bibnamefont {McCamey}}, \bibinfo {author} {\bibfnamefont {A.}~\bibnamefont
  {Laucht}},\ and\ \bibinfo {author} {\bibfnamefont {J.~J.}\ \bibnamefont
  {Pla}},\ }\bibfield  {title} {\bibinfo {title} {Room-temperature solid-state
  maser amplifier},\ }\href {https://doi.org/10.1103/PhysRevX.14.041066}
  {\bibfield  {journal} {\bibinfo  {journal} {Phys. Rev. X}\ }\textbf {\bibinfo
  {volume} {14}},\ \bibinfo {pages} {041066} (\bibinfo {year}
  {2024})}\BibitemShut {NoStop}%
\bibitem [{\citenamefont {Wu}\ \emph {et~al.}(2025)\citenamefont {Wu},
  \citenamefont {Han}, \citenamefont {Wang}, \citenamefont {Zheng},
  \citenamefont {Wang}, \citenamefont {Yang}, \citenamefont {Wang},
  \citenamefont {Zhang}, \citenamefont {Budker},\ and\ \citenamefont
  {Zhang}}]{wu2025detectingaxiondarkmatter}%
  \BibitemOpen
  \bibfield  {author} {\bibinfo {author} {\bibfnamefont {H.}~\bibnamefont
  {Wu}}, \bibinfo {author} {\bibfnamefont {Y.}~\bibnamefont {Han}}, \bibinfo
  {author} {\bibfnamefont {Z.}~\bibnamefont {Wang}}, \bibinfo {author}
  {\bibfnamefont {D.}~\bibnamefont {Zheng}}, \bibinfo {author} {\bibfnamefont
  {Y.}~\bibnamefont {Wang}}, \bibinfo {author} {\bibfnamefont {L.}~\bibnamefont
  {Yang}}, \bibinfo {author} {\bibfnamefont {Z.}~\bibnamefont {Wang}}, \bibinfo
  {author} {\bibfnamefont {B.}~\bibnamefont {Zhang}}, \bibinfo {author}
  {\bibfnamefont {D.}~\bibnamefont {Budker}},\ and\ \bibinfo {author}
  {\bibfnamefont {J.}~\bibnamefont {Zhang}},\ }\href
  {https://arxiv.org/abs/2512.17271} {\bibinfo {title} {Detecting axion dark
  matter with an organic molecular maser}} (\bibinfo {year} {2025}),\ \Eprint
  {https://arxiv.org/abs/2512.17271} {arXiv:2512.17271 [hep-ph]} \BibitemShut
  {NoStop}%
\bibitem [{\citenamefont {Wu}\ \emph {et~al.}(2022)\citenamefont {Wu},
  \citenamefont {Yang}, \citenamefont {Oxborrow}, \citenamefont {Jiang},
  \citenamefont {Zhao}, \citenamefont {Budker}, \citenamefont {Zhang},\ and\
  \citenamefont {Du}}]{doi:10.1126/sciadv.ade1613}%
  \BibitemOpen
  \bibfield  {author} {\bibinfo {author} {\bibfnamefont {H.}~\bibnamefont
  {Wu}}, \bibinfo {author} {\bibfnamefont {S.}~\bibnamefont {Yang}}, \bibinfo
  {author} {\bibfnamefont {M.}~\bibnamefont {Oxborrow}}, \bibinfo {author}
  {\bibfnamefont {M.}~\bibnamefont {Jiang}}, \bibinfo {author} {\bibfnamefont
  {Q.}~\bibnamefont {Zhao}}, \bibinfo {author} {\bibfnamefont {D.}~\bibnamefont
  {Budker}}, \bibinfo {author} {\bibfnamefont {B.}~\bibnamefont {Zhang}},\ and\
  \bibinfo {author} {\bibfnamefont {J.}~\bibnamefont {Du}},\ }\bibfield
  {title} {\bibinfo {title} {Enhanced quantum sensing with room-temperature
  solid-state masers},\ }\href {https://doi.org/10.1126/sciadv.ade1613}
  {\bibfield  {journal} {\bibinfo  {journal} {Science Advances}\ }\textbf
  {\bibinfo {volume} {8}},\ \bibinfo {pages} {eade1613} (\bibinfo {year}
  {2022})}\BibitemShut {NoStop}%
\bibitem [{\citenamefont {Breeze}\ \emph {et~al.}(2015)\citenamefont {Breeze},
  \citenamefont {Tan}, \citenamefont {Richards}, \citenamefont {Sathian},
  \citenamefont {Oxborrow},\ and\ \citenamefont {Alford}}]{Breeze2015}%
  \BibitemOpen
  \bibfield  {author} {\bibinfo {author} {\bibfnamefont {J.}~\bibnamefont
  {Breeze}}, \bibinfo {author} {\bibfnamefont {K.-J.}\ \bibnamefont {Tan}},
  \bibinfo {author} {\bibfnamefont {B.}~\bibnamefont {Richards}}, \bibinfo
  {author} {\bibfnamefont {J.}~\bibnamefont {Sathian}}, \bibinfo {author}
  {\bibfnamefont {M.}~\bibnamefont {Oxborrow}},\ and\ \bibinfo {author}
  {\bibfnamefont {N.~M.}\ \bibnamefont {Alford}},\ }\bibfield  {title}
  {\bibinfo {title} {Enhanced magnetic purcell effect in room-temperature
  masers},\ }\href {https://doi.org/10.1038/ncomms7215} {\bibfield  {journal}
  {\bibinfo  {journal} {Nature Communications}\ }\textbf {\bibinfo {volume}
  {6}},\ \bibinfo {pages} {6215} (\bibinfo {year} {2015})}\BibitemShut
  {NoStop}%
\bibitem [{\citenamefont {Miyanishi}\ \emph {et~al.}(2021)\citenamefont
  {Miyanishi}, \citenamefont {Segawa}, \citenamefont {Takeda}, \citenamefont
  {Ohki}, \citenamefont {Onoda}, \citenamefont {Ohshima}, \citenamefont {Abe},
  \citenamefont {Takashima}, \citenamefont {Takeuchi}, \citenamefont {Shames},
  \citenamefont {Morita}, \citenamefont {Wang}, \citenamefont {So},
  \citenamefont {Terada}, \citenamefont {Igarashi}, \citenamefont {Kagawa},
  \citenamefont {Kitagawa}, \citenamefont {Mizuochi}, \citenamefont
  {Shirakawa},\ and\ \citenamefont {Negoro}}]{mr-2-33-2021}%
  \BibitemOpen
  \bibfield  {author} {\bibinfo {author} {\bibfnamefont {K.}~\bibnamefont
  {Miyanishi}}, \bibinfo {author} {\bibfnamefont {T.~F.}\ \bibnamefont
  {Segawa}}, \bibinfo {author} {\bibfnamefont {K.}~\bibnamefont {Takeda}},
  \bibinfo {author} {\bibfnamefont {I.}~\bibnamefont {Ohki}}, \bibinfo {author}
  {\bibfnamefont {S.}~\bibnamefont {Onoda}}, \bibinfo {author} {\bibfnamefont
  {T.}~\bibnamefont {Ohshima}}, \bibinfo {author} {\bibfnamefont
  {H.}~\bibnamefont {Abe}}, \bibinfo {author} {\bibfnamefont {H.}~\bibnamefont
  {Takashima}}, \bibinfo {author} {\bibfnamefont {S.}~\bibnamefont {Takeuchi}},
  \bibinfo {author} {\bibfnamefont {A.~I.}\ \bibnamefont {Shames}}, \bibinfo
  {author} {\bibfnamefont {K.}~\bibnamefont {Morita}}, \bibinfo {author}
  {\bibfnamefont {Y.}~\bibnamefont {Wang}}, \bibinfo {author} {\bibfnamefont
  {F.~T.-K.}\ \bibnamefont {So}}, \bibinfo {author} {\bibfnamefont
  {D.}~\bibnamefont {Terada}}, \bibinfo {author} {\bibfnamefont
  {R.}~\bibnamefont {Igarashi}}, \bibinfo {author} {\bibfnamefont
  {A.}~\bibnamefont {Kagawa}}, \bibinfo {author} {\bibfnamefont
  {M.}~\bibnamefont {Kitagawa}}, \bibinfo {author} {\bibfnamefont
  {N.}~\bibnamefont {Mizuochi}}, \bibinfo {author} {\bibfnamefont
  {M.}~\bibnamefont {Shirakawa}},\ and\ \bibinfo {author} {\bibfnamefont
  {M.}~\bibnamefont {Negoro}},\ }\bibfield  {title} {\bibinfo {title}
  {Room-temperature hyperpolarization of polycrystalline samples with optically
  polarized triplet electrons: pentacene or nitrogen-vacancy center in
  diamond?},\ }\href {https://doi.org/10.5194/mr-2-33-2021} {\bibfield
  {journal} {\bibinfo  {journal} {Magnetic Resonance}\ }\textbf {\bibinfo
  {volume} {2}},\ \bibinfo {pages} {33} (\bibinfo {year} {2021})}\BibitemShut
  {NoStop}%
\bibitem [{\citenamefont {Bogatko}\ \emph {et~al.}(2016)\citenamefont
  {Bogatko}, \citenamefont {Haynes}, \citenamefont {Sathian}, \citenamefont
  {Wade}, \citenamefont {Kim}, \citenamefont {Tan}, \citenamefont {Breeze},
  \citenamefont {Salvadori}, \citenamefont {Horsfield},\ and\ \citenamefont
  {Oxborrow}}]{Bogatko2016}%
  \BibitemOpen
  \bibfield  {author} {\bibinfo {author} {\bibfnamefont {S.}~\bibnamefont
  {Bogatko}}, \bibinfo {author} {\bibfnamefont {P.~D.}\ \bibnamefont {Haynes}},
  \bibinfo {author} {\bibfnamefont {J.}~\bibnamefont {Sathian}}, \bibinfo
  {author} {\bibfnamefont {J.}~\bibnamefont {Wade}}, \bibinfo {author}
  {\bibfnamefont {J.-S.}\ \bibnamefont {Kim}}, \bibinfo {author} {\bibfnamefont
  {K.-J.}\ \bibnamefont {Tan}}, \bibinfo {author} {\bibfnamefont
  {J.}~\bibnamefont {Breeze}}, \bibinfo {author} {\bibfnamefont
  {E.}~\bibnamefont {Salvadori}}, \bibinfo {author} {\bibfnamefont
  {A.}~\bibnamefont {Horsfield}},\ and\ \bibinfo {author} {\bibfnamefont
  {M.}~\bibnamefont {Oxborrow}},\ }\bibfield  {title} {\bibinfo {title}
  {Molecular design of a room-temperature maser},\ }\href
  {https://doi.org/10.1021/acs.jpcc.6b00150} {\bibfield  {journal} {\bibinfo
  {journal} {The Journal of Physical Chemistry C}\ }\textbf {\bibinfo {volume}
  {120}},\ \bibinfo {pages} {8251} (\bibinfo {year} {2016})}\BibitemShut
  {NoStop}%
\bibitem [{\citenamefont {Ng}\ \emph {et~al.}(2023)\citenamefont {Ng},
  \citenamefont {Xu}, \citenamefont {Attwood}, \citenamefont {Wu},
  \citenamefont {Meng}, \citenamefont {Chen},\ and\ \citenamefont
  {Oxborrow}}]{adma.202300441}%
  \BibitemOpen
  \bibfield  {author} {\bibinfo {author} {\bibfnamefont {W.}~\bibnamefont
  {Ng}}, \bibinfo {author} {\bibfnamefont {X.}~\bibnamefont {Xu}}, \bibinfo
  {author} {\bibfnamefont {M.}~\bibnamefont {Attwood}}, \bibinfo {author}
  {\bibfnamefont {H.}~\bibnamefont {Wu}}, \bibinfo {author} {\bibfnamefont
  {Z.}~\bibnamefont {Meng}}, \bibinfo {author} {\bibfnamefont {X.}~\bibnamefont
  {Chen}},\ and\ \bibinfo {author} {\bibfnamefont {M.}~\bibnamefont
  {Oxborrow}},\ }\bibfield  {title} {\bibinfo {title} {Move aside pentacene:
  Diazapentacene-doped para-terphenyl, a zero-field room-temperature maser with
  strong coupling for cavity quantum electrodynamics},\ }\href
  {https://doi.org/https://doi.org/10.1002/adma.202300441} {\bibfield
  {journal} {\bibinfo  {journal} {Advanced Materials}\ }\textbf {\bibinfo
  {volume} {35}},\ \bibinfo {pages} {2300441} (\bibinfo {year}
  {2023})}\BibitemShut {NoStop}%
\bibitem [{\citenamefont {Schr{\"o}der}\ \emph {et~al.}(2022)\citenamefont
  {Schr{\"o}der}, \citenamefont {Rauber}, \citenamefont {Matt},\ and\
  \citenamefont {Kay}}]{Schroder2022}%
  \BibitemOpen
  \bibfield  {author} {\bibinfo {author} {\bibfnamefont {M.}~\bibnamefont
  {Schr{\"o}der}}, \bibinfo {author} {\bibfnamefont {D.}~\bibnamefont
  {Rauber}}, \bibinfo {author} {\bibfnamefont {C.}~\bibnamefont {Matt}},\ and\
  \bibinfo {author} {\bibfnamefont {C.~W.~M.}\ \bibnamefont {Kay}},\ }\bibfield
   {title} {\bibinfo {title} {Pentacene in 1,3,5-tri(1-naphtyl)benzene: A novel
  standard for transient epr spectroscopy at room temperature},\ }\href
  {https://doi.org/10.1007/s00723-021-01420-4} {\bibfield  {journal} {\bibinfo
  {journal} {Applied Magnetic Resonance}\ }\textbf {\bibinfo {volume} {53}},\
  \bibinfo {pages} {1043} (\bibinfo {year} {2022})}\BibitemShut {NoStop}%
\bibitem [{\citenamefont {KOUSKOV}\ \emph {et~al.}(1995)\citenamefont
  {KOUSKOV}, \citenamefont {SLOOP}, \citenamefont {LIU},\ and\ \citenamefont
  {LIN}}]{KOUSKOV19959}%
  \BibitemOpen
  \bibfield  {author} {\bibinfo {author} {\bibfnamefont {V.}~\bibnamefont
  {KOUSKOV}}, \bibinfo {author} {\bibfnamefont {D.~J.}\ \bibnamefont {SLOOP}},
  \bibinfo {author} {\bibfnamefont {S.-B.}\ \bibnamefont {LIU}},\ and\ \bibinfo
  {author} {\bibfnamefont {T.-S.}\ \bibnamefont {LIN}},\ }\bibfield  {title}
  {\bibinfo {title} {Pulsed transient nutation experiments of the photo-excited
  triplet state},\ }\href
  {https://doi.org/https://doi.org/10.1006/jmra.1995.9977} {\bibfield
  {journal} {\bibinfo  {journal} {Journal of Magnetic Resonance, Series A}\
  }\textbf {\bibinfo {volume} {117}},\ \bibinfo {pages} {9} (\bibinfo {year}
  {1995})}\BibitemShut {NoStop}%
\bibitem [{\citenamefont {K{\"o}hler}\ \emph {et~al.}(1993)\citenamefont
  {K{\"o}hler}, \citenamefont {Disselhorst}, \citenamefont {Donckers},
  \citenamefont {Groenen}, \citenamefont {Schmidt},\ and\ \citenamefont
  {Moerner}}]{Kohler1993}%
  \BibitemOpen
  \bibfield  {author} {\bibinfo {author} {\bibfnamefont {J.}~\bibnamefont
  {K{\"o}hler}}, \bibinfo {author} {\bibfnamefont {J.~A. J.~M.}\ \bibnamefont
  {Disselhorst}}, \bibinfo {author} {\bibfnamefont {M.~C. J.~M.}\ \bibnamefont
  {Donckers}}, \bibinfo {author} {\bibfnamefont {E.~J.~J.}\ \bibnamefont
  {Groenen}}, \bibinfo {author} {\bibfnamefont {J.}~\bibnamefont {Schmidt}},\
  and\ \bibinfo {author} {\bibfnamefont {W.~E.}\ \bibnamefont {Moerner}},\
  }\bibfield  {title} {\bibinfo {title} {Magnetic resonance of a single
  molecular spin},\ }\href {https://doi.org/10.1038/363242a0} {\bibfield
  {journal} {\bibinfo  {journal} {Nature}\ }\textbf {\bibinfo {volume} {363}},\
  \bibinfo {pages} {242} (\bibinfo {year} {1993})}\BibitemShut {NoStop}%
\bibitem [{\citenamefont {Attwood}\ \emph {et~al.}(2023)\citenamefont
  {Attwood}, \citenamefont {Xu}, \citenamefont {Newns}, \citenamefont {Meng},
  \citenamefont {Ingle}, \citenamefont {Wu}, \citenamefont {Chen},
  \citenamefont {Xu}, \citenamefont {Ng}, \citenamefont {Abiola}, \citenamefont
  {Stavros},\ and\ \citenamefont {Oxborrow}}]{Attwood2023}%
  \BibitemOpen
  \bibfield  {author} {\bibinfo {author} {\bibfnamefont {M.}~\bibnamefont
  {Attwood}}, \bibinfo {author} {\bibfnamefont {X.}~\bibnamefont {Xu}},
  \bibinfo {author} {\bibfnamefont {M.}~\bibnamefont {Newns}}, \bibinfo
  {author} {\bibfnamefont {Z.}~\bibnamefont {Meng}}, \bibinfo {author}
  {\bibfnamefont {R.~A.}\ \bibnamefont {Ingle}}, \bibinfo {author}
  {\bibfnamefont {H.}~\bibnamefont {Wu}}, \bibinfo {author} {\bibfnamefont
  {X.}~\bibnamefont {Chen}}, \bibinfo {author} {\bibfnamefont {W.}~\bibnamefont
  {Xu}}, \bibinfo {author} {\bibfnamefont {W.}~\bibnamefont {Ng}}, \bibinfo
  {author} {\bibfnamefont {T.~T.}\ \bibnamefont {Abiola}}, \bibinfo {author}
  {\bibfnamefont {V.~G.}\ \bibnamefont {Stavros}},\ and\ \bibinfo {author}
  {\bibfnamefont {M.}~\bibnamefont {Oxborrow}},\ }\bibfield  {title} {\bibinfo
  {title} {N-heteroacenes as an organic gain medium for room-temperature
  masers},\ }\href {https://doi.org/10.1021/acs.chemmater.3c00640} {\bibfield
  {journal} {\bibinfo  {journal} {Chemistry of Materials}\ }\textbf {\bibinfo
  {volume} {35}},\ \bibinfo {pages} {4498} (\bibinfo {year}
  {2023})}\BibitemShut {NoStop}%
\bibitem [{\citenamefont {Cui}\ \emph {et~al.}(2020)\citenamefont {Cui},
  \citenamefont {Liu}, \citenamefont {Li}, \citenamefont {Han}, \citenamefont
  {Ge}, \citenamefont {Zhang}, \citenamefont {Guo}, \citenamefont {Ye},\ and\
  \citenamefont {Tao}}]{Cui2020}%
  \BibitemOpen
  \bibfield  {author} {\bibinfo {author} {\bibfnamefont {S.}~\bibnamefont
  {Cui}}, \bibinfo {author} {\bibfnamefont {Y.}~\bibnamefont {Liu}}, \bibinfo
  {author} {\bibfnamefont {G.}~\bibnamefont {Li}}, \bibinfo {author}
  {\bibfnamefont {Q.}~\bibnamefont {Han}}, \bibinfo {author} {\bibfnamefont
  {C.}~\bibnamefont {Ge}}, \bibinfo {author} {\bibfnamefont {L.}~\bibnamefont
  {Zhang}}, \bibinfo {author} {\bibfnamefont {Q.}~\bibnamefont {Guo}}, \bibinfo
  {author} {\bibfnamefont {X.}~\bibnamefont {Ye}},\ and\ \bibinfo {author}
  {\bibfnamefont {X.}~\bibnamefont {Tao}},\ }\bibfield  {title} {\bibinfo
  {title} {Growth regulation of pentacene-doped p-terphenyl crystals on their
  physical properties for promising maser gain medium},\ }\href
  {https://doi.org/10.1021/acs.cgd.9b01190} {\bibfield  {journal} {\bibinfo
  {journal} {Crystal Growth {\&} Design}\ }\textbf {\bibinfo {volume} {20}},\
  \bibinfo {pages} {783} (\bibinfo {year} {2020})}\BibitemShut {NoStop}%
\bibitem [{\citenamefont {Daniels}(1996)}]{doi:10.1049/ecej:19960402}%
  \BibitemOpen
  \bibfield  {author} {\bibinfo {author} {\bibfnamefont {D.}~\bibnamefont
  {Daniels}},\ }\bibfield  {title} {\bibinfo {title} {Surface-penetrating
  radar},\ }\href {https://doi.org/10.1049/ecej:19960402} {\bibfield  {journal}
  {\bibinfo  {journal} {Electronics \& Communication Engineering Journal}\
  }\textbf {\bibinfo {volume} {8}},\ \bibinfo {pages} {165} (\bibinfo {year}
  {1996})}\BibitemShut {NoStop}%
\bibitem [{\citenamefont {Griffiths}\ \emph {et~al.}(2015)\citenamefont
  {Griffiths}, \citenamefont {Cohen}, \citenamefont {Watts}, \citenamefont
  {Mokole}, \citenamefont {Baker}, \citenamefont {Wicks},\ and\ \citenamefont
  {Blunt}}]{6967722}%
  \BibitemOpen
  \bibfield  {author} {\bibinfo {author} {\bibfnamefont {H.}~\bibnamefont
  {Griffiths}}, \bibinfo {author} {\bibfnamefont {L.}~\bibnamefont {Cohen}},
  \bibinfo {author} {\bibfnamefont {S.}~\bibnamefont {Watts}}, \bibinfo
  {author} {\bibfnamefont {E.}~\bibnamefont {Mokole}}, \bibinfo {author}
  {\bibfnamefont {C.}~\bibnamefont {Baker}}, \bibinfo {author} {\bibfnamefont
  {M.}~\bibnamefont {Wicks}},\ and\ \bibinfo {author} {\bibfnamefont
  {S.}~\bibnamefont {Blunt}},\ }\bibfield  {title} {\bibinfo {title} {Radar
  spectrum engineering and management: Technical and regulatory issues},\
  }\href {https://doi.org/10.1109/JPROC.2014.2365517} {\bibfield  {journal}
  {\bibinfo  {journal} {Proceedings of the IEEE}\ }\textbf {\bibinfo {volume}
  {103}},\ \bibinfo {pages} {85} (\bibinfo {year} {2015})}\BibitemShut
  {NoStop}%
\bibitem [{\citenamefont {{van Haarlem, M. P.}}\ \emph
  {et~al.}(2013)\citenamefont {{van Haarlem, M. P.}}, \citenamefont {{Wise, M.
  W.}}, \citenamefont {{Gunst, A. W.}}, \citenamefont {{Heald, G.}},
  \citenamefont {{McKean, J. P.}}, \citenamefont {{Hessels, J. W. T.}},
  \citenamefont {{de Bruyn, A. G.}}, \citenamefont {{Nijboer, R.}},
  \citenamefont {{Swinbank, J.}}, \citenamefont {{Fallows, R.}}, \citenamefont
  {{Brentjens, M.}}, \citenamefont {{Nelles, A.}}, \citenamefont {{Beck, R.}},
  \citenamefont {{Falcke, H.}}, \citenamefont {{Fender, R.}}, \citenamefont
  {{Hörandel, J.}}, \citenamefont {{Koopmans, L. V. E.}}, \citenamefont
  {{Mann, G.}}, \citenamefont {{Miley, G.}}, \citenamefont {{Röttgering, H.}},
  \citenamefont {{Stappers, B. W.}}, \citenamefont {{Wijers, R. A. M. J.}},
  \citenamefont {{Zaroubi, S.}}, \citenamefont {{van den Akker, M.}},
  \citenamefont {{Alexov, A.}}, \citenamefont {{Anderson, J.}}, \citenamefont
  {{Anderson, K.}}, \citenamefont {{van Ardenne, A.}}, \citenamefont {{Arts,
  M.}}, \citenamefont {{Asgekar, A.}}, \citenamefont {{Avruch, I. M.}},
  \citenamefont {{Batejat, F.}}, \citenamefont {{Bähren, L.}}, \citenamefont
  {{Bell, M. E.}}, \citenamefont {{Bell, M. R.}}, \citenamefont {{van Bemmel,
  I.}}, \citenamefont {{Bennema, P.}}, \citenamefont {{Bentum, M. J.}},
  \citenamefont {{Bernardi, G.}}, \citenamefont {{Best, P.}}, \citenamefont
  {{Bîrzan, L.}}, \citenamefont {{Bonafede, A.}}, \citenamefont {{Boonstra,
  A.-J.}}, \citenamefont {{Braun, R.}}, \citenamefont {{Bregman, J.}},
  \citenamefont {{Breitling, F.}}, \citenamefont {{van de Brink, R. H.}},
  \citenamefont {{Broderick, J.}}, \citenamefont {{Broekema, P. C.}},
  \citenamefont {{Brouw, W. N.}}, \citenamefont {{Brüggen, M.}}, \citenamefont
  {{Butcher, H. R.}}, \citenamefont {{van Cappellen, W.}}, \citenamefont
  {{Ciardi, B.}}, \citenamefont {{Coenen, T.}}, \citenamefont {{Conway, J.}},
  \citenamefont {{Coolen, A.}}, \citenamefont {{Corstanje, A.}}, \citenamefont
  {{Damstra, S.}}, \citenamefont {{Davies, O.}}, \citenamefont {{Deller, A.
  T.}}, \citenamefont {{Dettmar, R.-J.}}, \citenamefont {{van Diepen, G.}},
  \citenamefont {{Dijkstra, K.}}, \citenamefont {{Donker, P.}}, \citenamefont
  {{Doorduin, A.}}, \citenamefont {{Dromer, J.}}, \citenamefont {{Drost, M.}},
  \citenamefont {{van Duin, A.}}, \citenamefont {{Eislöffel, J.}},
  \citenamefont {{van Enst, J.}}, \citenamefont {{Ferrari, C.}}, \citenamefont
  {{Frieswijk, W.}}, \citenamefont {{Gankema, H.}}, \citenamefont {{Garrett, M.
  A.}}, \citenamefont {{de Gasperin, F.}}, \citenamefont {{Gerbers, M.}},
  \citenamefont {{de Geus, E.}}, \citenamefont {{Grießmeier, J.-M.}},
  \citenamefont {{Grit, T.}}, \citenamefont {{Gruppen, P.}}, \citenamefont
  {{Hamaker, J. P.}}, \citenamefont {{Hassall, T.}}, \citenamefont {{Hoeft,
  M.}}, \citenamefont {{Holties, H. A.}}, \citenamefont {{Horneffer, A.}},
  \citenamefont {{van der Horst, A.}}, \citenamefont {{van Houwelingen, A.}},
  \citenamefont {{Huijgen, A.}}, \citenamefont {{Iacobelli, M.}}, \citenamefont
  {{Intema, H.}}, \citenamefont {{Jackson, N.}}, \citenamefont {{Jelic, V.}},
  \citenamefont {{de Jong, A.}}, \citenamefont {{Juette, E.}}, \citenamefont
  {{Kant, D.}}, \citenamefont {{Karastergiou, A.}}, \citenamefont {{Koers,
  A.}}, \citenamefont {{Kollen, H.}}, \citenamefont {{Kondratiev, V. I.}},
  \citenamefont {{Kooistra, E.}}, \citenamefont {{Koopman, Y.}}, \citenamefont
  {{Koster, A.}}, \citenamefont {{Kuniyoshi, M.}}, \citenamefont {{Kramer,
  M.}}, \citenamefont {{Kuper, G.}}, \citenamefont {{Lambropoulos, P.}},
  \citenamefont {{Law, C.}}, \citenamefont {{van Leeuwen, J.}}, \citenamefont
  {{Lemaitre, J.}}, \citenamefont {{Loose, M.}}, \citenamefont {{Maat, P.}},
  \citenamefont {{Macario, G.}}, \citenamefont {{Markoff, S.}}, \citenamefont
  {{Masters, J.}}, \citenamefont {{McFadden, R. A.}}, \citenamefont
  {{McKay-Bukowski, D.}}, \citenamefont {{Meijering, H.}}, \citenamefont
  {{Meulman, H.}}, \citenamefont {{Mevius, M.}}, \citenamefont {{Middelberg,
  E.}}, \citenamefont {{Millenaar, R.}}, \citenamefont {{Miller-Jones, J. C.
  A.}}, \citenamefont {{Mohan, R. N.}}, \citenamefont {{Mol, J. D.}},
  \citenamefont {{Morawietz, J.}}, \citenamefont {{Morganti, R.}},
  \citenamefont {{Mulcahy, D. D.}}, \citenamefont {{Mulder, E.}}, \citenamefont
  {{Munk, H.}}, \citenamefont {{Nieuwenhuis, L.}}, \citenamefont {{van
  Nieuwpoort, R.}}, \citenamefont {{Noordam, J. E.}}, \citenamefont {{Norden,
  M.}}, \citenamefont {{Noutsos, A.}}, \citenamefont {{Offringa, A. R.}},
  \citenamefont {{Olofsson, H.}}, \citenamefont {{Omar, A.}}, \citenamefont
  {{Orrú, E.}}, \citenamefont {{Overeem, R.}}, \citenamefont {{Paas, H.}},
  \citenamefont {{Pandey-Pommier, M.}}, \citenamefont {{Pandey, V. N.}},
  \citenamefont {{Pizzo, R.}}, \citenamefont {{Polatidis, A.}}, \citenamefont
  {{Rafferty, D.}}, \citenamefont {{Rawlings, S.}}, \citenamefont {{Reich,
  W.}}, \citenamefont {{de Reijer, J.-P.}}, \citenamefont {{Reitsma, J.}},
  \citenamefont {{Renting, G. A.}}, \citenamefont {{Riemers, P.}},
  \citenamefont {{Rol, E.}}, \citenamefont {{Romein, J. W.}}, \citenamefont
  {{Roosjen, J.}}, \citenamefont {{Ruiter, M.}}, \citenamefont {{Scaife, A.}},
  \citenamefont {{van der Schaaf, K.}}, \citenamefont {{Scheers, B.}},
  \citenamefont {{Schellart, P.}}, \citenamefont {{Schoenmakers, A.}},
  \citenamefont {{Schoonderbeek, G.}}, \citenamefont {{Serylak, M.}},
  \citenamefont {{Shulevski, A.}}, \citenamefont {{Sluman, J.}}, \citenamefont
  {{Smirnov, O.}}, \citenamefont {{Sobey, C.}}, \citenamefont {{Spreeuw, H.}},
  \citenamefont {{Steinmetz, M.}}, \citenamefont {{Sterks, C. G. M.}},
  \citenamefont {{Stiepel, H.-J.}}, \citenamefont {{Stuurwold, K.}},
  \citenamefont {{Tagger, M.}}, \citenamefont {{Tang, Y.}}, \citenamefont
  {{Tasse, C.}}, \citenamefont {{Thomas, I.}}, \citenamefont {{Thoudam, S.}},
  \citenamefont {{Toribio, M. C.}}, \citenamefont {{van der Tol, B.}},
  \citenamefont {{Usov, O.}}, \citenamefont {{van Veelen, M.}}, \citenamefont
  {{van der Veen, A.-J.}}, \citenamefont {{ter Veen, S.}}, \citenamefont
  {{Verbiest, J. P. W.}}, \citenamefont {{Vermeulen, R.}}, \citenamefont
  {{Vermaas, N.}}, \citenamefont {{Vocks, C.}}, \citenamefont {{Vogt, C.}},
  \citenamefont {{de Vos, M.}}, \citenamefont {{van der Wal, E.}},
  \citenamefont {{van Weeren, R.}}, \citenamefont {{Weggemans, H.}},
  \citenamefont {{Weltevrede, P.}}, \citenamefont {{White, S.}}, \citenamefont
  {{Wijnholds, S. J.}}, \citenamefont {{Wilhelmsson, T.}}, \citenamefont
  {{Wucknitz, O.}}, \citenamefont {{Yatawatta, S.}}, \citenamefont {{Zarka,
  P.}}, \citenamefont {{Zensus, A.}},\ and\ \citenamefont {{van Zwieten,
  J.}}}]{refId0}%
  \BibitemOpen
  \bibfield  {author} {\bibinfo {author} {\bibnamefont {{van Haarlem, M. P.}}},
  \bibinfo {author} {\bibnamefont {{Wise, M. W.}}}, \bibinfo {author}
  {\bibnamefont {{Gunst, A. W.}}}, \bibinfo {author} {\bibnamefont {{Heald,
  G.}}}, \bibinfo {author} {\bibnamefont {{McKean, J. P.}}}, \bibinfo {author}
  {\bibnamefont {{Hessels, J. W. T.}}}, \bibinfo {author} {\bibnamefont {{de
  Bruyn, A. G.}}}, \bibinfo {author} {\bibnamefont {{Nijboer, R.}}}, \bibinfo
  {author} {\bibnamefont {{Swinbank, J.}}}, \bibinfo {author} {\bibnamefont
  {{Fallows, R.}}}, \bibinfo {author} {\bibnamefont {{Brentjens, M.}}},
  \bibinfo {author} {\bibnamefont {{Nelles, A.}}}, \bibinfo {author}
  {\bibnamefont {{Beck, R.}}}, \bibinfo {author} {\bibnamefont {{Falcke, H.}}},
  \bibinfo {author} {\bibnamefont {{Fender, R.}}}, \bibinfo {author}
  {\bibnamefont {{Hörandel, J.}}}, \bibinfo {author} {\bibnamefont {{Koopmans,
  L. V. E.}}}, \bibinfo {author} {\bibnamefont {{Mann, G.}}}, \bibinfo {author}
  {\bibnamefont {{Miley, G.}}}, \bibinfo {author} {\bibnamefont {{Röttgering,
  H.}}}, \bibinfo {author} {\bibnamefont {{Stappers, B. W.}}}, \bibinfo
  {author} {\bibnamefont {{Wijers, R. A. M. J.}}}, \bibinfo {author}
  {\bibnamefont {{Zaroubi, S.}}}, \bibinfo {author} {\bibnamefont {{van den
  Akker, M.}}}, \bibinfo {author} {\bibnamefont {{Alexov, A.}}}, \bibinfo
  {author} {\bibnamefont {{Anderson, J.}}}, \bibinfo {author} {\bibnamefont
  {{Anderson, K.}}}, \bibinfo {author} {\bibnamefont {{van Ardenne, A.}}},
  \bibinfo {author} {\bibnamefont {{Arts, M.}}}, \bibinfo {author}
  {\bibnamefont {{Asgekar, A.}}}, \bibinfo {author} {\bibnamefont {{Avruch, I.
  M.}}}, \bibinfo {author} {\bibnamefont {{Batejat, F.}}}, \bibinfo {author}
  {\bibnamefont {{Bähren, L.}}}, \bibinfo {author} {\bibnamefont {{Bell, M.
  E.}}}, \bibinfo {author} {\bibnamefont {{Bell, M. R.}}}, \bibinfo {author}
  {\bibnamefont {{van Bemmel, I.}}}, \bibinfo {author} {\bibnamefont {{Bennema,
  P.}}}, \bibinfo {author} {\bibnamefont {{Bentum, M. J.}}}, \bibinfo {author}
  {\bibnamefont {{Bernardi, G.}}}, \bibinfo {author} {\bibnamefont {{Best,
  P.}}}, \bibinfo {author} {\bibnamefont {{Bîrzan, L.}}}, \bibinfo {author}
  {\bibnamefont {{Bonafede, A.}}}, \bibinfo {author} {\bibnamefont {{Boonstra,
  A.-J.}}}, \bibinfo {author} {\bibnamefont {{Braun, R.}}}, \bibinfo {author}
  {\bibnamefont {{Bregman, J.}}}, \bibinfo {author} {\bibnamefont {{Breitling,
  F.}}}, \bibinfo {author} {\bibnamefont {{van de Brink, R. H.}}}, \bibinfo
  {author} {\bibnamefont {{Broderick, J.}}}, \bibinfo {author} {\bibnamefont
  {{Broekema, P. C.}}}, \bibinfo {author} {\bibnamefont {{Brouw, W. N.}}},
  \bibinfo {author} {\bibnamefont {{Brüggen, M.}}}, \bibinfo {author}
  {\bibnamefont {{Butcher, H. R.}}}, \bibinfo {author} {\bibnamefont {{van
  Cappellen, W.}}}, \bibinfo {author} {\bibnamefont {{Ciardi, B.}}}, \bibinfo
  {author} {\bibnamefont {{Coenen, T.}}}, \bibinfo {author} {\bibnamefont
  {{Conway, J.}}}, \bibinfo {author} {\bibnamefont {{Coolen, A.}}}, \bibinfo
  {author} {\bibnamefont {{Corstanje, A.}}}, \bibinfo {author} {\bibnamefont
  {{Damstra, S.}}}, \bibinfo {author} {\bibnamefont {{Davies, O.}}}, \bibinfo
  {author} {\bibnamefont {{Deller, A. T.}}}, \bibinfo {author} {\bibnamefont
  {{Dettmar, R.-J.}}}, \bibinfo {author} {\bibnamefont {{van Diepen, G.}}},
  \bibinfo {author} {\bibnamefont {{Dijkstra, K.}}}, \bibinfo {author}
  {\bibnamefont {{Donker, P.}}}, \bibinfo {author} {\bibnamefont {{Doorduin,
  A.}}}, \bibinfo {author} {\bibnamefont {{Dromer, J.}}}, \bibinfo {author}
  {\bibnamefont {{Drost, M.}}}, \bibinfo {author} {\bibnamefont {{van Duin,
  A.}}}, \bibinfo {author} {\bibnamefont {{Eislöffel, J.}}}, \bibinfo {author}
  {\bibnamefont {{van Enst, J.}}}, \bibinfo {author} {\bibnamefont {{Ferrari,
  C.}}}, \bibinfo {author} {\bibnamefont {{Frieswijk, W.}}}, \bibinfo {author}
  {\bibnamefont {{Gankema, H.}}}, \bibinfo {author} {\bibnamefont {{Garrett, M.
  A.}}}, \bibinfo {author} {\bibnamefont {{de Gasperin, F.}}}, \bibinfo
  {author} {\bibnamefont {{Gerbers, M.}}}, \bibinfo {author} {\bibnamefont {{de
  Geus, E.}}}, \bibinfo {author} {\bibnamefont {{Grießmeier, J.-M.}}},
  \bibinfo {author} {\bibnamefont {{Grit, T.}}}, \bibinfo {author}
  {\bibnamefont {{Gruppen, P.}}}, \bibinfo {author} {\bibnamefont {{Hamaker, J.
  P.}}}, \bibinfo {author} {\bibnamefont {{Hassall, T.}}}, \bibinfo {author}
  {\bibnamefont {{Hoeft, M.}}}, \bibinfo {author} {\bibnamefont {{Holties, H.
  A.}}}, \bibinfo {author} {\bibnamefont {{Horneffer, A.}}}, \bibinfo {author}
  {\bibnamefont {{van der Horst, A.}}}, \bibinfo {author} {\bibnamefont {{van
  Houwelingen, A.}}}, \bibinfo {author} {\bibnamefont {{Huijgen, A.}}},
  \bibinfo {author} {\bibnamefont {{Iacobelli, M.}}}, \bibinfo {author}
  {\bibnamefont {{Intema, H.}}}, \bibinfo {author} {\bibnamefont {{Jackson,
  N.}}}, \bibinfo {author} {\bibnamefont {{Jelic, V.}}}, \bibinfo {author}
  {\bibnamefont {{de Jong, A.}}}, \bibinfo {author} {\bibnamefont {{Juette,
  E.}}}, \bibinfo {author} {\bibnamefont {{Kant, D.}}}, \bibinfo {author}
  {\bibnamefont {{Karastergiou, A.}}}, \bibinfo {author} {\bibnamefont {{Koers,
  A.}}}, \bibinfo {author} {\bibnamefont {{Kollen, H.}}}, \bibinfo {author}
  {\bibnamefont {{Kondratiev, V. I.}}}, \bibinfo {author} {\bibnamefont
  {{Kooistra, E.}}}, \bibinfo {author} {\bibnamefont {{Koopman, Y.}}}, \bibinfo
  {author} {\bibnamefont {{Koster, A.}}}, \bibinfo {author} {\bibnamefont
  {{Kuniyoshi, M.}}}, \bibinfo {author} {\bibnamefont {{Kramer, M.}}}, \bibinfo
  {author} {\bibnamefont {{Kuper, G.}}}, \bibinfo {author} {\bibnamefont
  {{Lambropoulos, P.}}}, \bibinfo {author} {\bibnamefont {{Law, C.}}}, \bibinfo
  {author} {\bibnamefont {{van Leeuwen, J.}}}, \bibinfo {author} {\bibnamefont
  {{Lemaitre, J.}}}, \bibinfo {author} {\bibnamefont {{Loose, M.}}}, \bibinfo
  {author} {\bibnamefont {{Maat, P.}}}, \bibinfo {author} {\bibnamefont
  {{Macario, G.}}}, \bibinfo {author} {\bibnamefont {{Markoff, S.}}}, \bibinfo
  {author} {\bibnamefont {{Masters, J.}}}, \bibinfo {author} {\bibnamefont
  {{McFadden, R. A.}}}, \bibinfo {author} {\bibnamefont {{McKay-Bukowski,
  D.}}}, \bibinfo {author} {\bibnamefont {{Meijering, H.}}}, \bibinfo {author}
  {\bibnamefont {{Meulman, H.}}}, \bibinfo {author} {\bibnamefont {{Mevius,
  M.}}}, \bibinfo {author} {\bibnamefont {{Middelberg, E.}}}, \bibinfo {author}
  {\bibnamefont {{Millenaar, R.}}}, \bibinfo {author} {\bibnamefont
  {{Miller-Jones, J. C. A.}}}, \bibinfo {author} {\bibnamefont {{Mohan, R.
  N.}}}, \bibinfo {author} {\bibnamefont {{Mol, J. D.}}}, \bibinfo {author}
  {\bibnamefont {{Morawietz, J.}}}, \bibinfo {author} {\bibnamefont {{Morganti,
  R.}}}, \bibinfo {author} {\bibnamefont {{Mulcahy, D. D.}}}, \bibinfo {author}
  {\bibnamefont {{Mulder, E.}}}, \bibinfo {author} {\bibnamefont {{Munk, H.}}},
  \bibinfo {author} {\bibnamefont {{Nieuwenhuis, L.}}}, \bibinfo {author}
  {\bibnamefont {{van Nieuwpoort, R.}}}, \bibinfo {author} {\bibnamefont
  {{Noordam, J. E.}}}, \bibinfo {author} {\bibnamefont {{Norden, M.}}},
  \bibinfo {author} {\bibnamefont {{Noutsos, A.}}}, \bibinfo {author}
  {\bibnamefont {{Offringa, A. R.}}}, \bibinfo {author} {\bibnamefont
  {{Olofsson, H.}}}, \bibinfo {author} {\bibnamefont {{Omar, A.}}}, \bibinfo
  {author} {\bibnamefont {{Orrú, E.}}}, \bibinfo {author} {\bibnamefont
  {{Overeem, R.}}}, \bibinfo {author} {\bibnamefont {{Paas, H.}}}, \bibinfo
  {author} {\bibnamefont {{Pandey-Pommier, M.}}}, \bibinfo {author}
  {\bibnamefont {{Pandey, V. N.}}}, \bibinfo {author} {\bibnamefont {{Pizzo,
  R.}}}, \bibinfo {author} {\bibnamefont {{Polatidis, A.}}}, \bibinfo {author}
  {\bibnamefont {{Rafferty, D.}}}, \bibinfo {author} {\bibnamefont {{Rawlings,
  S.}}}, \bibinfo {author} {\bibnamefont {{Reich, W.}}}, \bibinfo {author}
  {\bibnamefont {{de Reijer, J.-P.}}}, \bibinfo {author} {\bibnamefont
  {{Reitsma, J.}}}, \bibinfo {author} {\bibnamefont {{Renting, G. A.}}},
  \bibinfo {author} {\bibnamefont {{Riemers, P.}}}, \bibinfo {author}
  {\bibnamefont {{Rol, E.}}}, \bibinfo {author} {\bibnamefont {{Romein, J.
  W.}}}, \bibinfo {author} {\bibnamefont {{Roosjen, J.}}}, \bibinfo {author}
  {\bibnamefont {{Ruiter, M.}}}, \bibinfo {author} {\bibnamefont {{Scaife,
  A.}}}, \bibinfo {author} {\bibnamefont {{van der Schaaf, K.}}}, \bibinfo
  {author} {\bibnamefont {{Scheers, B.}}}, \bibinfo {author} {\bibnamefont
  {{Schellart, P.}}}, \bibinfo {author} {\bibnamefont {{Schoenmakers, A.}}},
  \bibinfo {author} {\bibnamefont {{Schoonderbeek, G.}}}, \bibinfo {author}
  {\bibnamefont {{Serylak, M.}}}, \bibinfo {author} {\bibnamefont {{Shulevski,
  A.}}}, \bibinfo {author} {\bibnamefont {{Sluman, J.}}}, \bibinfo {author}
  {\bibnamefont {{Smirnov, O.}}}, \bibinfo {author} {\bibnamefont {{Sobey,
  C.}}}, \bibinfo {author} {\bibnamefont {{Spreeuw, H.}}}, \bibinfo {author}
  {\bibnamefont {{Steinmetz, M.}}}, \bibinfo {author} {\bibnamefont {{Sterks,
  C. G. M.}}}, \bibinfo {author} {\bibnamefont {{Stiepel, H.-J.}}}, \bibinfo
  {author} {\bibnamefont {{Stuurwold, K.}}}, \bibinfo {author} {\bibnamefont
  {{Tagger, M.}}}, \bibinfo {author} {\bibnamefont {{Tang, Y.}}}, \bibinfo
  {author} {\bibnamefont {{Tasse, C.}}}, \bibinfo {author} {\bibnamefont
  {{Thomas, I.}}}, \bibinfo {author} {\bibnamefont {{Thoudam, S.}}}, \bibinfo
  {author} {\bibnamefont {{Toribio, M. C.}}}, \bibinfo {author} {\bibnamefont
  {{van der Tol, B.}}}, \bibinfo {author} {\bibnamefont {{Usov, O.}}}, \bibinfo
  {author} {\bibnamefont {{van Veelen, M.}}}, \bibinfo {author} {\bibnamefont
  {{van der Veen, A.-J.}}}, \bibinfo {author} {\bibnamefont {{ter Veen, S.}}},
  \bibinfo {author} {\bibnamefont {{Verbiest, J. P. W.}}}, \bibinfo {author}
  {\bibnamefont {{Vermeulen, R.}}}, \bibinfo {author} {\bibnamefont {{Vermaas,
  N.}}}, \bibinfo {author} {\bibnamefont {{Vocks, C.}}}, \bibinfo {author}
  {\bibnamefont {{Vogt, C.}}}, \bibinfo {author} {\bibnamefont {{de Vos, M.}}},
  \bibinfo {author} {\bibnamefont {{van der Wal, E.}}}, \bibinfo {author}
  {\bibnamefont {{van Weeren, R.}}}, \bibinfo {author} {\bibnamefont
  {{Weggemans, H.}}}, \bibinfo {author} {\bibnamefont {{Weltevrede, P.}}},
  \bibinfo {author} {\bibnamefont {{White, S.}}}, \bibinfo {author}
  {\bibnamefont {{Wijnholds, S. J.}}}, \bibinfo {author} {\bibnamefont
  {{Wilhelmsson, T.}}}, \bibinfo {author} {\bibnamefont {{Wucknitz, O.}}},
  \bibinfo {author} {\bibnamefont {{Yatawatta, S.}}}, \bibinfo {author}
  {\bibnamefont {{Zarka, P.}}}, \bibinfo {author} {\bibnamefont {{Zensus,
  A.}}},\ and\ \bibinfo {author} {\bibnamefont {{van Zwieten, J.}}},\
  }\bibfield  {title} {\bibinfo {title} {Lofar: The low-frequency array},\
  }\href {https://doi.org/10.1051/0004-6361/201220873} {\bibfield  {journal}
  {\bibinfo  {journal} {A\,\& A}\ }\textbf {\bibinfo {volume} {556}},\ \bibinfo
  {pages} {A2} (\bibinfo {year} {2013})}\BibitemShut {NoStop}%
\bibitem [{\citenamefont {Takeda}\ \emph {et~al.}(2002)\citenamefont {Takeda},
  \citenamefont {Takegoshi},\ and\ \citenamefont {Terao}}]{10.1063/1.1499124}%
  \BibitemOpen
  \bibfield  {author} {\bibinfo {author} {\bibfnamefont {K.}~\bibnamefont
  {Takeda}}, \bibinfo {author} {\bibfnamefont {K.}~\bibnamefont {Takegoshi}},\
  and\ \bibinfo {author} {\bibfnamefont {T.}~\bibnamefont {Terao}},\ }\bibfield
   {title} {\bibinfo {title} {Zero-field electron spin resonance and
  theoretical studies of light penetration into single crystal and
  polycrystalline material doped with molecules photoexcitable to the triplet
  state via intersystem crossing},\ }\href {https://doi.org/10.1063/1.1499124}
  {\bibfield  {journal} {\bibinfo  {journal} {The Journal of Chemical Physics}\
  }\textbf {\bibinfo {volume} {117}},\ \bibinfo {pages} {4940} (\bibinfo {year}
  {2002})}\BibitemShut {NoStop}%
\bibitem [{\citenamefont {Sloop}\ \emph {et~al.}(1981)\citenamefont {Sloop},
  \citenamefont {Yu}, \citenamefont {Lin},\ and\ \citenamefont
  {Weissman}}]{10.1063/1.442520}%
  \BibitemOpen
  \bibfield  {author} {\bibinfo {author} {\bibfnamefont {D.~J.}\ \bibnamefont
  {Sloop}}, \bibinfo {author} {\bibfnamefont {H.}~\bibnamefont {Yu}}, \bibinfo
  {author} {\bibfnamefont {T.}~\bibnamefont {Lin}},\ and\ \bibinfo {author}
  {\bibfnamefont {S.~I.}\ \bibnamefont {Weissman}},\ }\bibfield  {title}
  {\bibinfo {title} {Electron spin echoes of a photoexcited triplet: Pentacene
  in p‐terphenyl crystals},\ }\href {https://doi.org/10.1063/1.442520}
  {\bibfield  {journal} {\bibinfo  {journal} {The Journal of Chemical Physics}\
  }\textbf {\bibinfo {volume} {75}},\ \bibinfo {pages} {3746} (\bibinfo {year}
  {1981})}\BibitemShut {NoStop}%
\bibitem [{\citenamefont {Yang}\ \emph {et~al.}(2000)\citenamefont {Yang},
  \citenamefont {Sloop}, \citenamefont {Weissman},\ and\ \citenamefont
  {Lin}}]{10.1063/1.1326069}%
  \BibitemOpen
  \bibfield  {author} {\bibinfo {author} {\bibfnamefont {T.-C.}\ \bibnamefont
  {Yang}}, \bibinfo {author} {\bibfnamefont {D.~J.}\ \bibnamefont {Sloop}},
  \bibinfo {author} {\bibfnamefont {S.~I.}\ \bibnamefont {Weissman}},\ and\
  \bibinfo {author} {\bibfnamefont {T.-S.}\ \bibnamefont {Lin}},\ }\bibfield
  {title} {\bibinfo {title} {Zero-field magnetic resonance of the photo-excited
  triplet state of pentacene at room temperature},\ }\href
  {https://doi.org/10.1063/1.1326069} {\bibfield  {journal} {\bibinfo
  {journal} {The Journal of Chemical Physics}\ }\textbf {\bibinfo {volume}
  {113}},\ \bibinfo {pages} {11194} (\bibinfo {year} {2000})}\BibitemShut
  {NoStop}%
\bibitem [{\citenamefont {Wu}\ \emph {et~al.}(2019)\citenamefont {Wu},
  \citenamefont {Ng}, \citenamefont {Mirkhanov}, \citenamefont {Amirzhan},
  \citenamefont {Nitnara},\ and\ \citenamefont {Oxborrow}}]{Wu2019}%
  \BibitemOpen
  \bibfield  {author} {\bibinfo {author} {\bibfnamefont {H.}~\bibnamefont
  {Wu}}, \bibinfo {author} {\bibfnamefont {W.}~\bibnamefont {Ng}}, \bibinfo
  {author} {\bibfnamefont {S.}~\bibnamefont {Mirkhanov}}, \bibinfo {author}
  {\bibfnamefont {A.}~\bibnamefont {Amirzhan}}, \bibinfo {author}
  {\bibfnamefont {S.}~\bibnamefont {Nitnara}},\ and\ \bibinfo {author}
  {\bibfnamefont {M.}~\bibnamefont {Oxborrow}},\ }\bibfield  {title} {\bibinfo
  {title} {Unraveling the room-temperature spin dynamics of photoexcited
  pentacene in its lowest triplet state at zero field},\ }\href
  {https://doi.org/10.1021/acs.jpcc.9b08439} {\bibfield  {journal} {\bibinfo
  {journal} {The Journal of Physical Chemistry C}\ }\textbf {\bibinfo {volume}
  {123}},\ \bibinfo {pages} {24275} (\bibinfo {year} {2019})}\BibitemShut
  {NoStop}%
\bibitem [{\citenamefont {Ng}\ \emph {et~al.}(2021)\citenamefont {Ng},
  \citenamefont {Zhang}, \citenamefont {Wu}, \citenamefont {Nevjestic},
  \citenamefont {White},\ and\ \citenamefont {Oxborrow}}]{Ng2021}%
  \BibitemOpen
  \bibfield  {author} {\bibinfo {author} {\bibfnamefont {W.}~\bibnamefont
  {Ng}}, \bibinfo {author} {\bibfnamefont {S.}~\bibnamefont {Zhang}}, \bibinfo
  {author} {\bibfnamefont {H.}~\bibnamefont {Wu}}, \bibinfo {author}
  {\bibfnamefont {I.}~\bibnamefont {Nevjestic}}, \bibinfo {author}
  {\bibfnamefont {A.~J.~P.}\ \bibnamefont {White}},\ and\ \bibinfo {author}
  {\bibfnamefont {M.}~\bibnamefont {Oxborrow}},\ }\bibfield  {title} {\bibinfo
  {title} {Exploring the triplet spin dynamics of the charge-transfer
  co-crystal phenazine/1,2,4,5-tetracyanobenzene for potential use in organic
  maser gain media},\ }\href {https://doi.org/10.1021/acs.jpcc.1c01654}
  {\bibfield  {journal} {\bibinfo  {journal} {The Journal of Physical Chemistry
  C}\ }\textbf {\bibinfo {volume} {125}},\ \bibinfo {pages} {14718} (\bibinfo
  {year} {2021})}\BibitemShut {NoStop}%
\bibitem [{\citenamefont {Yao}\ \emph {et~al.}(2017)\citenamefont {Yao},
  \citenamefont {Gui}, \citenamefont {Rao}, \citenamefont {Kaur}, \citenamefont
  {Chen}, \citenamefont {Lu}, \citenamefont {Xiao}, \citenamefont {Guo},
  \citenamefont {Marzlin},\ and\ \citenamefont {Hu}}]{Yao2017}%
  \BibitemOpen
  \bibfield  {author} {\bibinfo {author} {\bibfnamefont {B.}~\bibnamefont
  {Yao}}, \bibinfo {author} {\bibfnamefont {Y.~S.}\ \bibnamefont {Gui}},
  \bibinfo {author} {\bibfnamefont {J.~W.}\ \bibnamefont {Rao}}, \bibinfo
  {author} {\bibfnamefont {S.}~\bibnamefont {Kaur}}, \bibinfo {author}
  {\bibfnamefont {X.~S.}\ \bibnamefont {Chen}}, \bibinfo {author}
  {\bibfnamefont {W.}~\bibnamefont {Lu}}, \bibinfo {author} {\bibfnamefont
  {Y.}~\bibnamefont {Xiao}}, \bibinfo {author} {\bibfnamefont {H.}~\bibnamefont
  {Guo}}, \bibinfo {author} {\bibfnamefont {K.-P.}\ \bibnamefont {Marzlin}},\
  and\ \bibinfo {author} {\bibfnamefont {C.-M.}\ \bibnamefont {Hu}},\
  }\bibfield  {title} {\bibinfo {title} {Cooperative polariton dynamics in
  feedback-coupled cavities},\ }\href
  {https://doi.org/10.1038/s41467-017-01796-7} {\bibfield  {journal} {\bibinfo
  {journal} {Nature Communications}\ }\textbf {\bibinfo {volume} {8}},\
  \bibinfo {pages} {1437} (\bibinfo {year} {2017})}\BibitemShut {NoStop}%
\bibitem [{\citenamefont {Yao}\ \emph {et~al.}(2023)\citenamefont {Yao},
  \citenamefont {Gui}, \citenamefont {Rao}, \citenamefont {Zhang},
  \citenamefont {Lu},\ and\ \citenamefont {Hu}}]{PhysRevLett.130.146702}%
  \BibitemOpen
  \bibfield  {author} {\bibinfo {author} {\bibfnamefont {B.}~\bibnamefont
  {Yao}}, \bibinfo {author} {\bibfnamefont {Y.~S.}\ \bibnamefont {Gui}},
  \bibinfo {author} {\bibfnamefont {J.~W.}\ \bibnamefont {Rao}}, \bibinfo
  {author} {\bibfnamefont {Y.~H.}\ \bibnamefont {Zhang}}, \bibinfo {author}
  {\bibfnamefont {W.}~\bibnamefont {Lu}},\ and\ \bibinfo {author}
  {\bibfnamefont {C.-M.}\ \bibnamefont {Hu}},\ }\bibfield  {title} {\bibinfo
  {title} {Coherent microwave emission of gain-driven polaritons},\ }\href
  {https://doi.org/10.1103/PhysRevLett.130.146702} {\bibfield  {journal}
  {\bibinfo  {journal} {Phys. Rev. Lett.}\ }\textbf {\bibinfo {volume} {130}},\
  \bibinfo {pages} {146702} (\bibinfo {year} {2023})}\BibitemShut {NoStop}%
\bibitem [{\citenamefont {Jiang}\ \emph {et~al.}(2021)\citenamefont {Jiang},
  \citenamefont {Su}, \citenamefont {Wu}, \citenamefont {Peng},\ and\
  \citenamefont {Budker}}]{doi:10.1126/sciadv.abe0719}%
  \BibitemOpen
  \bibfield  {author} {\bibinfo {author} {\bibfnamefont {M.}~\bibnamefont
  {Jiang}}, \bibinfo {author} {\bibfnamefont {H.}~\bibnamefont {Su}}, \bibinfo
  {author} {\bibfnamefont {Z.}~\bibnamefont {Wu}}, \bibinfo {author}
  {\bibfnamefont {X.}~\bibnamefont {Peng}},\ and\ \bibinfo {author}
  {\bibfnamefont {D.}~\bibnamefont {Budker}},\ }\bibfield  {title} {\bibinfo
  {title} {Floquet maser},\ }\href {https://doi.org/10.1126/sciadv.abe0719}
  {\bibfield  {journal} {\bibinfo  {journal} {Science Advances}\ }\textbf
  {\bibinfo {volume} {7}},\ \bibinfo {pages} {eabe0719} (\bibinfo {year}
  {2021})}\BibitemShut {NoStop}%
\bibitem [{\citenamefont {Xiang}\ \emph {et~al.}(2025)\citenamefont {Xiang},
  \citenamefont {Bugnon}, \citenamefont {Khatibi~Moghaddam},\ and\
  \citenamefont {Fleury}}]{Xiang2025}%
  \BibitemOpen
  \bibfield  {author} {\bibinfo {author} {\bibfnamefont {R.}~\bibnamefont
  {Xiang}}, \bibinfo {author} {\bibfnamefont {P.}~\bibnamefont {Bugnon}},
  \bibinfo {author} {\bibfnamefont {M.}~\bibnamefont {Khatibi~Moghaddam}},\
  and\ \bibinfo {author} {\bibfnamefont {R.}~\bibnamefont {Fleury}},\
  }\bibfield  {title} {\bibinfo {title} {All-metallic magnetic purcell
  enhancement in a thermally stable room-temperature maser},\ }\href
  {https://doi.org/10.1038/s41467-025-66016-z} {\bibfield  {journal} {\bibinfo
  {journal} {Nature Communications}\ }\textbf {\bibinfo {volume} {16}},\
  \bibinfo {pages} {11214} (\bibinfo {year} {2025})}\BibitemShut {NoStop}%
\bibitem [{\citenamefont {Geyer}\ \emph {et~al.}(2005)\citenamefont {Geyer},
  \citenamefont {Riddle}, \citenamefont {Krupka},\ and\ \citenamefont
  {Boatner}}]{DielectricProperties}%
  \BibitemOpen
  \bibfield  {author} {\bibinfo {author} {\bibfnamefont {R.}~\bibnamefont
  {Geyer}}, \bibinfo {author} {\bibfnamefont {B.}~\bibnamefont {Riddle}},
  \bibinfo {author} {\bibfnamefont {J.}~\bibnamefont {Krupka}},\ and\ \bibinfo
  {author} {\bibfnamefont {L.}~\bibnamefont {Boatner}},\ }\bibfield  {title}
  {\bibinfo {title} {Microwave dielectric properties of single-crystal quantum
  paraelectrics ktao3 and srtio3 at cryogenic temperatures},\ }\href
  {https://doi.org/10.1063/1.1905789} {\bibfield  {journal} {\bibinfo
  {journal} {Journal of Applied Physics}\ }\textbf {\bibinfo {volume} {97}},\
  \bibinfo {pages} {104111} (\bibinfo {year} {2005})}\BibitemShut {NoStop}%
\bibitem [{\citenamefont {Zollitsch}\ \emph {et~al.}(2023)\citenamefont
  {Zollitsch}, \citenamefont {Ruloff}, \citenamefont {Fett}, \citenamefont
  {Wiedemann}, \citenamefont {Richter}, \citenamefont {Breeze},\ and\
  \citenamefont {Kay}}]{Zollitsch2023}%
  \BibitemOpen
  \bibfield  {author} {\bibinfo {author} {\bibfnamefont {C.~W.}\ \bibnamefont
  {Zollitsch}}, \bibinfo {author} {\bibfnamefont {S.}~\bibnamefont {Ruloff}},
  \bibinfo {author} {\bibfnamefont {Y.}~\bibnamefont {Fett}}, \bibinfo {author}
  {\bibfnamefont {H.~T.~A.}\ \bibnamefont {Wiedemann}}, \bibinfo {author}
  {\bibfnamefont {R.}~\bibnamefont {Richter}}, \bibinfo {author} {\bibfnamefont
  {J.~D.}\ \bibnamefont {Breeze}},\ and\ \bibinfo {author} {\bibfnamefont
  {C.~W.~M.}\ \bibnamefont {Kay}},\ }\bibfield  {title} {\bibinfo {title}
  {Maser threshold characterization by resonator q-factor tuning},\ }\href
  {https://doi.org/10.1038/s42005-023-01418-3} {\bibfield  {journal} {\bibinfo
  {journal} {Communications Physics}\ }\textbf {\bibinfo {volume} {6}},\
  \bibinfo {pages} {295} (\bibinfo {year} {2023})}\BibitemShut {NoStop}%
\bibitem [{\citenamefont {Poole}(1983)}]{poole1983electron}%
  \BibitemOpen
  \bibfield  {author} {\bibinfo {author} {\bibfnamefont {C.}~\bibnamefont
  {Poole}},\ }\href {https://books.google.co.jp/books?id=7v-nQgAACAAJ} {\emph
  {\bibinfo {title} {Electron Spin Resonance: A Comprehensive Treatise on
  Experimental Techniques}}},\ A Wiley-Interscience publication\ (\bibinfo
  {publisher} {Wiley},\ \bibinfo {year} {1983})\BibitemShut {NoStop}%
\bibitem [{\citenamefont {Wang}\ \emph {et~al.}(2022)\citenamefont {Wang},
  \citenamefont {Kong}, \citenamefont {Zhao}, \citenamefont {Huang},
  \citenamefont {Yu}, \citenamefont {Wang}, \citenamefont {Shi},\ and\
  \citenamefont {Du}}]{doi:10.1126/sciadv.abq8158}%
  \BibitemOpen
  \bibfield  {author} {\bibinfo {author} {\bibfnamefont {Z.}~\bibnamefont
  {Wang}}, \bibinfo {author} {\bibfnamefont {F.}~\bibnamefont {Kong}}, \bibinfo
  {author} {\bibfnamefont {P.}~\bibnamefont {Zhao}}, \bibinfo {author}
  {\bibfnamefont {Z.}~\bibnamefont {Huang}}, \bibinfo {author} {\bibfnamefont
  {P.}~\bibnamefont {Yu}}, \bibinfo {author} {\bibfnamefont {Y.}~\bibnamefont
  {Wang}}, \bibinfo {author} {\bibfnamefont {F.}~\bibnamefont {Shi}},\ and\
  \bibinfo {author} {\bibfnamefont {J.}~\bibnamefont {Du}},\ }\bibfield
  {title} {\bibinfo {title} {Picotesla magnetometry of microwave fields with
  diamond sensors},\ }\href {https://doi.org/10.1126/sciadv.abq8158} {\bibfield
   {journal} {\bibinfo  {journal} {Science Advances}\ }\textbf {\bibinfo
  {volume} {8}},\ \bibinfo {pages} {eabq8158} (\bibinfo {year}
  {2022})}\BibitemShut {NoStop}%
\bibitem [{\citenamefont {Meinel}\ \emph {et~al.}(2021)\citenamefont {Meinel},
  \citenamefont {Vorobyov}, \citenamefont {Yavkin}, \citenamefont {Dasari},
  \citenamefont {Sumiya}, \citenamefont {Onoda}, \citenamefont {Isoya},\ and\
  \citenamefont {Wrachtrup}}]{Meinel2021}%
  \BibitemOpen
  \bibfield  {author} {\bibinfo {author} {\bibfnamefont {J.}~\bibnamefont
  {Meinel}}, \bibinfo {author} {\bibfnamefont {V.}~\bibnamefont {Vorobyov}},
  \bibinfo {author} {\bibfnamefont {B.}~\bibnamefont {Yavkin}}, \bibinfo
  {author} {\bibfnamefont {D.}~\bibnamefont {Dasari}}, \bibinfo {author}
  {\bibfnamefont {H.}~\bibnamefont {Sumiya}}, \bibinfo {author} {\bibfnamefont
  {S.}~\bibnamefont {Onoda}}, \bibinfo {author} {\bibfnamefont
  {J.}~\bibnamefont {Isoya}},\ and\ \bibinfo {author} {\bibfnamefont
  {J.}~\bibnamefont {Wrachtrup}},\ }\bibfield  {title} {\bibinfo {title}
  {Heterodyne sensing of microwaves with a quantum sensor},\ }\href
  {https://doi.org/10.1038/s41467-021-22714-y} {\bibfield  {journal} {\bibinfo
  {journal} {Nature Communications}\ }\textbf {\bibinfo {volume} {12}},\
  \bibinfo {pages} {2737} (\bibinfo {year} {2021})}\BibitemShut {NoStop}%
\bibitem [{\citenamefont {Alsid}\ \emph {et~al.}(2023)\citenamefont {Alsid},
  \citenamefont {Schloss}, \citenamefont {Steinecker}, \citenamefont {Barry},
  \citenamefont {Maccabe}, \citenamefont {Wang}, \citenamefont {Cappellaro},\
  and\ \citenamefont {Braje}}]{PhysRevApplied.19.054095}%
  \BibitemOpen
  \bibfield  {author} {\bibinfo {author} {\bibfnamefont {S.~T.}\ \bibnamefont
  {Alsid}}, \bibinfo {author} {\bibfnamefont {J.~M.}\ \bibnamefont {Schloss}},
  \bibinfo {author} {\bibfnamefont {M.~H.}\ \bibnamefont {Steinecker}},
  \bibinfo {author} {\bibfnamefont {J.~F.}\ \bibnamefont {Barry}}, \bibinfo
  {author} {\bibfnamefont {A.~C.}\ \bibnamefont {Maccabe}}, \bibinfo {author}
  {\bibfnamefont {G.}~\bibnamefont {Wang}}, \bibinfo {author} {\bibfnamefont
  {P.}~\bibnamefont {Cappellaro}},\ and\ \bibinfo {author} {\bibfnamefont
  {D.~A.}\ \bibnamefont {Braje}},\ }\bibfield  {title} {\bibinfo {title}
  {Solid-state microwave magnetometer with picotesla-level sensitivity},\
  }\href {https://doi.org/10.1103/PhysRevApplied.19.054095} {\bibfield
  {journal} {\bibinfo  {journal} {Phys. Rev. Appl.}\ }\textbf {\bibinfo
  {volume} {19}},\ \bibinfo {pages} {054095} (\bibinfo {year}
  {2023})}\BibitemShut {NoStop}%
\bibitem [{\citenamefont {Wang}\ \emph {et~al.}(2015)\citenamefont {Wang},
  \citenamefont {Yuan}, \citenamefont {Huang}, \citenamefont {Rong},
  \citenamefont {Wang}, \citenamefont {Xu}, \citenamefont {Duan}, \citenamefont
  {Ju}, \citenamefont {Shi},\ and\ \citenamefont {Du}}]{Wang2015}%
  \BibitemOpen
  \bibfield  {author} {\bibinfo {author} {\bibfnamefont {P.}~\bibnamefont
  {Wang}}, \bibinfo {author} {\bibfnamefont {Z.}~\bibnamefont {Yuan}}, \bibinfo
  {author} {\bibfnamefont {P.}~\bibnamefont {Huang}}, \bibinfo {author}
  {\bibfnamefont {X.}~\bibnamefont {Rong}}, \bibinfo {author} {\bibfnamefont
  {M.}~\bibnamefont {Wang}}, \bibinfo {author} {\bibfnamefont {X.}~\bibnamefont
  {Xu}}, \bibinfo {author} {\bibfnamefont {C.}~\bibnamefont {Duan}}, \bibinfo
  {author} {\bibfnamefont {C.}~\bibnamefont {Ju}}, \bibinfo {author}
  {\bibfnamefont {F.}~\bibnamefont {Shi}},\ and\ \bibinfo {author}
  {\bibfnamefont {J.}~\bibnamefont {Du}},\ }\bibfield  {title} {\bibinfo
  {title} {High-resolution vector microwave magnetometry based on solid-state
  spins in diamond},\ }\href {https://doi.org/10.1038/ncomms7631} {\bibfield
  {journal} {\bibinfo  {journal} {Nature Communications}\ }\textbf {\bibinfo
  {volume} {6}},\ \bibinfo {pages} {6631} (\bibinfo {year} {2015})}\BibitemShut
  {NoStop}%
\bibitem [{\citenamefont {Chen}\ \emph {et~al.}(2023)\citenamefont {Chen},
  \citenamefont {Wang}, \citenamefont {Shan}, \citenamefont {Zhang},
  \citenamefont {Feng}, \citenamefont {Zheng}, \citenamefont {Dong},
  \citenamefont {Guo},\ and\ \citenamefont {Sun}}]{Chen2023}%
  \BibitemOpen
  \bibfield  {author} {\bibinfo {author} {\bibfnamefont {X.-D.}\ \bibnamefont
  {Chen}}, \bibinfo {author} {\bibfnamefont {E.-H.}\ \bibnamefont {Wang}},
  \bibinfo {author} {\bibfnamefont {L.-K.}\ \bibnamefont {Shan}}, \bibinfo
  {author} {\bibfnamefont {S.-C.}\ \bibnamefont {Zhang}}, \bibinfo {author}
  {\bibfnamefont {C.}~\bibnamefont {Feng}}, \bibinfo {author} {\bibfnamefont
  {Y.}~\bibnamefont {Zheng}}, \bibinfo {author} {\bibfnamefont
  {Y.}~\bibnamefont {Dong}}, \bibinfo {author} {\bibfnamefont {G.-C.}\
  \bibnamefont {Guo}},\ and\ \bibinfo {author} {\bibfnamefont {F.-W.}\
  \bibnamefont {Sun}},\ }\bibfield  {title} {\bibinfo {title} {Quantum enhanced
  radio detection and ranging with solid spins},\ }\href
  {https://doi.org/10.1038/s41467-023-36929-8} {\bibfield  {journal} {\bibinfo
  {journal} {Nature Communications}\ }\textbf {\bibinfo {volume} {14}},\
  \bibinfo {pages} {1288} (\bibinfo {year} {2023})}\BibitemShut {NoStop}%
\bibitem [{\citenamefont {Chen}\ \emph {et~al.}(2019)\citenamefont {Chen},
  \citenamefont {Li}, \citenamefont {Zheng}, \citenamefont {Li}, \citenamefont
  {Du}, \citenamefont {Dong}, \citenamefont {Dong}, \citenamefont {Guo},\ and\
  \citenamefont {Sun}}]{PhysRevApplied.12.044039}%
  \BibitemOpen
  \bibfield  {author} {\bibinfo {author} {\bibfnamefont {X.-D.}\ \bibnamefont
  {Chen}}, \bibinfo {author} {\bibfnamefont {D.-F.}\ \bibnamefont {Li}},
  \bibinfo {author} {\bibfnamefont {Y.}~\bibnamefont {Zheng}}, \bibinfo
  {author} {\bibfnamefont {S.}~\bibnamefont {Li}}, \bibinfo {author}
  {\bibfnamefont {B.}~\bibnamefont {Du}}, \bibinfo {author} {\bibfnamefont
  {Y.}~\bibnamefont {Dong}}, \bibinfo {author} {\bibfnamefont {C.-H.}\
  \bibnamefont {Dong}}, \bibinfo {author} {\bibfnamefont {G.-C.}\ \bibnamefont
  {Guo}},\ and\ \bibinfo {author} {\bibfnamefont {F.-W.}\ \bibnamefont {Sun}},\
  }\bibfield  {title} {\bibinfo {title} {Superresolution multifunctional
  sensing with the nitrogen-vacancy center in diamond},\ }\href
  {https://doi.org/10.1103/PhysRevApplied.12.044039} {\bibfield  {journal}
  {\bibinfo  {journal} {Phys. Rev. Appl.}\ }\textbf {\bibinfo {volume} {12}},\
  \bibinfo {pages} {044039} (\bibinfo {year} {2019})}\BibitemShut {NoStop}%
\bibitem [{\citenamefont {Eisenach}\ \emph {et~al.}(2021)\citenamefont
  {Eisenach}, \citenamefont {Barry}, \citenamefont {O'Keeffe}, \citenamefont
  {Schloss}, \citenamefont {Steinecker}, \citenamefont {Englund},\ and\
  \citenamefont {Braje}}]{Eisenach2021}%
  \BibitemOpen
  \bibfield  {author} {\bibinfo {author} {\bibfnamefont {E.~R.}\ \bibnamefont
  {Eisenach}}, \bibinfo {author} {\bibfnamefont {J.~F.}\ \bibnamefont {Barry}},
  \bibinfo {author} {\bibfnamefont {M.~F.}\ \bibnamefont {O'Keeffe}}, \bibinfo
  {author} {\bibfnamefont {J.~M.}\ \bibnamefont {Schloss}}, \bibinfo {author}
  {\bibfnamefont {M.~H.}\ \bibnamefont {Steinecker}}, \bibinfo {author}
  {\bibfnamefont {D.~R.}\ \bibnamefont {Englund}},\ and\ \bibinfo {author}
  {\bibfnamefont {D.~A.}\ \bibnamefont {Braje}},\ }\bibfield  {title} {\bibinfo
  {title} {Cavity-enhanced microwave readout of a solid-state spin sensor},\
  }\href {https://doi.org/10.1038/s41467-021-21256-7} {\bibfield  {journal}
  {\bibinfo  {journal} {Nature Communications}\ }\textbf {\bibinfo {volume}
  {12}},\ \bibinfo {pages} {1357} (\bibinfo {year} {2021})}\BibitemShut
  {NoStop}%
\bibitem [{\citenamefont {Pravdivtsev}\ \emph {et~al.}(2020)\citenamefont
  {Pravdivtsev}, \citenamefont {Sönnichsen},\ and\ \citenamefont
  {Hövener}}]{https://doi.org/10.1002/cphc.201901056}%
  \BibitemOpen
  \bibfield  {author} {\bibinfo {author} {\bibfnamefont {A.~N.}\ \bibnamefont
  {Pravdivtsev}}, \bibinfo {author} {\bibfnamefont {F.~D.}\ \bibnamefont
  {Sönnichsen}},\ and\ \bibinfo {author} {\bibfnamefont {J.-B.}\ \bibnamefont
  {Hövener}},\ }\bibfield  {title} {\bibinfo {title} {Continuous radio
  amplification by stimulated emission of radiation using parahydrogen induced
  polarization (phip-raser) at 14 tesla},\ }\href
  {https://doi.org/https://doi.org/10.1002/cphc.201901056} {\bibfield
  {journal} {\bibinfo  {journal} {ChemPhysChem}\ }\textbf {\bibinfo {volume}
  {21}},\ \bibinfo {pages} {667} (\bibinfo {year} {2020})}\BibitemShut
  {NoStop}%
\bibitem [{\citenamefont {Suefke}\ \emph {et~al.}(2017)\citenamefont {Suefke},
  \citenamefont {Lehmkuhl}, \citenamefont {Liebisch}, \citenamefont
  {Bl{\"u}mich},\ and\ \citenamefont {Appelt}}]{Suefke2017}%
  \BibitemOpen
  \bibfield  {author} {\bibinfo {author} {\bibfnamefont {M.}~\bibnamefont
  {Suefke}}, \bibinfo {author} {\bibfnamefont {S.}~\bibnamefont {Lehmkuhl}},
  \bibinfo {author} {\bibfnamefont {A.}~\bibnamefont {Liebisch}}, \bibinfo
  {author} {\bibfnamefont {B.}~\bibnamefont {Bl{\"u}mich}},\ and\ \bibinfo
  {author} {\bibfnamefont {S.}~\bibnamefont {Appelt}},\ }\bibfield  {title}
  {\bibinfo {title} {Para-hydrogen raser delivers sub-millihertz resolution in
  nuclear magnetic resonance},\ }\href {https://doi.org/10.1038/nphys4076}
  {\bibfield  {journal} {\bibinfo  {journal} {Nature Physics}\ }\textbf
  {\bibinfo {volume} {13}},\ \bibinfo {pages} {568} (\bibinfo {year}
  {2017})}\BibitemShut {NoStop}%
\bibitem [{\citenamefont {Lehmkuhl}\ \emph {et~al.}(2026)\citenamefont
  {Lehmkuhl}, \citenamefont {Fleischer}, \citenamefont {Yang}, \citenamefont
  {Chekmenev}, \citenamefont {Theis}, \citenamefont {Appelt}, \citenamefont
  {Korvink},\ and\ \citenamefont
  {Jouda}}]{https://doi.org/10.1002/anie.202525699}%
  \BibitemOpen
  \bibfield  {author} {\bibinfo {author} {\bibfnamefont {S.}~\bibnamefont
  {Lehmkuhl}}, \bibinfo {author} {\bibfnamefont {S.}~\bibnamefont {Fleischer}},
  \bibinfo {author} {\bibfnamefont {J.}~\bibnamefont {Yang}}, \bibinfo {author}
  {\bibfnamefont {E.~Y.}\ \bibnamefont {Chekmenev}}, \bibinfo {author}
  {\bibfnamefont {T.}~\bibnamefont {Theis}}, \bibinfo {author} {\bibfnamefont
  {S.}~\bibnamefont {Appelt}}, \bibinfo {author} {\bibfnamefont {J.~G.}\
  \bibnamefont {Korvink}},\ and\ \bibinfo {author} {\bibfnamefont
  {M.}~\bibnamefont {Jouda}},\ }\bibfield  {title} {\bibinfo {title} {Rapid
  raser mri},\ }\href {https://doi.org/https://doi.org/10.1002/anie.202525699}
  {\bibfield  {journal} {\bibinfo  {journal} {Angewandte Chemie International
  Edition}\ }\textbf {\bibinfo {volume} {65}},\ \bibinfo {pages} {e25699}
  (\bibinfo {year} {2026})}\BibitemShut {NoStop}%
\bibitem [{\citenamefont {Chamberlain}\ \emph {et~al.}(2007)\citenamefont
  {Chamberlain}, \citenamefont {Park}, \citenamefont {Corum}, \citenamefont
  {Yacoub}, \citenamefont {Ugurbil}, \citenamefont {Jack~Jr},\ and\
  \citenamefont {Garwood}}]{https://doi.org/10.1002/mrm.21396}%
  \BibitemOpen
  \bibfield  {author} {\bibinfo {author} {\bibfnamefont {R.}~\bibnamefont
  {Chamberlain}}, \bibinfo {author} {\bibfnamefont {J.-Y.}\ \bibnamefont
  {Park}}, \bibinfo {author} {\bibfnamefont {C.}~\bibnamefont {Corum}},
  \bibinfo {author} {\bibfnamefont {E.}~\bibnamefont {Yacoub}}, \bibinfo
  {author} {\bibfnamefont {K.}~\bibnamefont {Ugurbil}}, \bibinfo {author}
  {\bibfnamefont {C.~R.}\ \bibnamefont {Jack~Jr}},\ and\ \bibinfo {author}
  {\bibfnamefont {M.}~\bibnamefont {Garwood}},\ }\bibfield  {title} {\bibinfo
  {title} {Raser: A new ultrafast magnetic resonance imaging method},\ }\href
  {https://doi.org/https://doi.org/10.1002/mrm.21396} {\bibfield  {journal}
  {\bibinfo  {journal} {Magnetic Resonance in Medicine}\ }\textbf {\bibinfo
  {volume} {58}},\ \bibinfo {pages} {794} (\bibinfo {year} {2007})}\BibitemShut
  {NoStop}%
\bibitem [{\citenamefont {Ng}\ \emph {et~al.}(2024)\citenamefont {Ng},
  \citenamefont {Wen}, \citenamefont {Attwood}, \citenamefont {Jones},
  \citenamefont {Oxborrow}, \citenamefont {Alford},\ and\ \citenamefont
  {Arroo}}]{10.1063/5.0181318}%
  \BibitemOpen
  \bibfield  {author} {\bibinfo {author} {\bibfnamefont {W.}~\bibnamefont
  {Ng}}, \bibinfo {author} {\bibfnamefont {Y.}~\bibnamefont {Wen}}, \bibinfo
  {author} {\bibfnamefont {M.}~\bibnamefont {Attwood}}, \bibinfo {author}
  {\bibfnamefont {D.~C.}\ \bibnamefont {Jones}}, \bibinfo {author}
  {\bibfnamefont {M.}~\bibnamefont {Oxborrow}}, \bibinfo {author}
  {\bibfnamefont {N.~M.}\ \bibnamefont {Alford}},\ and\ \bibinfo {author}
  {\bibfnamefont {D.~M.}\ \bibnamefont {Arroo}},\ }\bibfield  {title} {\bibinfo
  {title} {“maser-in-a-shoebox”: A portable plug-and-play maser device at
  room temperature and zero magnetic field},\ }\href
  {https://doi.org/10.1063/5.0181318} {\bibfield  {journal} {\bibinfo
  {journal} {Applied Physics Letters}\ }\textbf {\bibinfo {volume} {124}},\
  \bibinfo {pages} {044004} (\bibinfo {year} {2024})}\BibitemShut {NoStop}%
\bibitem [{\citenamefont {Ng}\ \emph {et~al.}(2025)\citenamefont {Ng},
  \citenamefont {Wen}, \citenamefont {Alford},\ and\ \citenamefont
  {Arroo}}]{PhysRevApplied.23.054064}%
  \BibitemOpen
  \bibfield  {author} {\bibinfo {author} {\bibfnamefont {W.}~\bibnamefont
  {Ng}}, \bibinfo {author} {\bibfnamefont {Y.}~\bibnamefont {Wen}}, \bibinfo
  {author} {\bibfnamefont {N.~M.}\ \bibnamefont {Alford}},\ and\ \bibinfo
  {author} {\bibfnamefont {D.~M.}\ \bibnamefont {Arroo}},\ }\bibfield  {title}
  {\bibinfo {title} {Portable maser oscillator at room temperature with reduced
  magnetic field requirements through spatial orientation},\ }\href
  {https://doi.org/10.1103/PhysRevApplied.23.054064} {\bibfield  {journal}
  {\bibinfo  {journal} {Phys. Rev. Appl.}\ }\textbf {\bibinfo {volume} {23}},\
  \bibinfo {pages} {054064} (\bibinfo {year} {2025})}\BibitemShut {NoStop}%
\end{thebibliography}

%

\end{document}